\documentclass[aps,10pt]{revtex4}

\usepackage{amsmath,amssymb,bm,epsfig}
\usepackage{color}
\usepackage{natbib}
\usepackage{hyperref} 
\usepackage{ulem}
\usepackage{graphicx}
\usepackage{xcolor}
\usepackage{pifont}
\usepackage{amsfonts}
\usepackage{dsfont,mathrsfs}
\usepackage{cancel}

\newcommand{\nn}{\nonumber}

\newcommand{\MB}[1]{\left|#1\right|}
\newcommand{\FB}[1]{\left(#1\right)}
\newcommand{\SB}[1]{\left\{#1\right\}}
\newcommand{\TB}[1]{\left[#1\right]}

\newcommand{\scrL}{\mathscr{L}}

\newcommand{\munu}{{\mu\nu}}
\newcommand{\alphabeta}{{\alpha\beta}}

\newcommand{\RE}{\text{Re}}
\newcommand{\Tr}{\text{Tr}}

\newcommand{\del}{\partial}
\newcommand{\deltilde}{\tilde{\partial}}

\def\be {\begin{equation}}
\def\ee {\end{equation}}
\def\nn {\nonumber}
\def\bea {\begin{eqnarray}}
\def\eea {\end{eqnarray}}
\def\op {\hat{O}}

\def\ps{p\!\!\!/}
\def\us{u\!\!\!/}
\def\bs{b\!\!\!/}
\def\m{m_f}

\def\p{\partial}
\def\e{\epsilon}
\def\g{\gamma}
\def\gf{\gamma_5}

\def\s{\sigma}

\def\k{\kappa}

\begin{document}
\title{Effect of external magnetic field on nucleon mass in hot and dense medium : Inverse Magnetic Catalysis in Walecka Model }
\author{Arghya Mukherjee$^{a,c}$}
\email{arghya.mukherjee@saha.ac.in}
\author{Snigdha Ghosh$^{b,c,d}$}
\email{snigdha.physics@gmail.com, snigdha.ghosh@iitgn.ac.in}
\author{Mahatsab Mandal$^e$}
\email{mahatsab@gmail.com}
\author{Sourav Sarkar$^{b,c}$}
\email{sourav@vecc.gov.in}
\author{ Pradip Roy$^{a,c}$}
\email{pradipk.roy@saha.ac.in}
\affiliation{$^a$Saha Institute of Nuclear Physics, 1/AF Bidhannagar, Kolkata - 700064, India}
\affiliation{$^b$Variable Energy Cyclotron Centre, 1/AF Bidhannagar, Kolkata 700 064, India}
\affiliation{$^c$Homi Bhabha National Institute, Training School Complex, Anushaktinagar, Mumbai - 400085, India}
\affiliation{$^d$Indian Institute of Technology Gandhinagar, Palaj, Gandhinagar 382355, Gujarat, India}
\affiliation{$^e$Government General Degree College at Kalna-I, Burdwan, West Bengal - 713405, India}
\begin{abstract}
Vacuum to nuclear matter phase transition has been studied in presence of constant external background magnetic field with the  mean field approximation in Walecka model.
The anomalous nucleon magnetic moment has been taken into account using the modified  ``weak'' field expansion of the  fermion propagator
having non-trivial correction terms for charged as well as for   neutral particles. 
The effect of nucleon magnetic moment is found to favour  the magnetic catalysis effect at zero temperature and zero baryon density.  However, extending  the 
study to finite temperatures, it is observed that the anomalous nuclear  magnetic moment plays a crucial role in characterizing the qualitative 
behaviour of vacuum to nuclear matter phase transition even in case of the weak external magnetic fields .  The critical 
temperature corresponding to   the vacuum to nuclear medium phase transition is observed to  decrease with 
the external magnetic field which can  be identified as the inverse magnetic catalysis  in Walecka model whereas  the opposite behaviour is obtained in case of 
vanishing  magnetic moment indicating  magnetic catalysis.  
\end{abstract}
\maketitle
%
\section{Introduction}
Understanding Quantum Chromodynamics (QCD) in presence of magnetic background has gained lots of contemporary research interests \cite{ln871}. 
It is important to study QCD in presence of 
external magnetic field not only  for  its  relevance with the astrophysical phenomena \cite{prep442,prl95,prd76,prl100,prdN76,prl105,prd82}
but also due to  the possibility of strong magnetic field generation in 
non-central heavy-ion collision \cite{ijmp} which sets the stage for investigation of this  magnetic modifications. Although the background fields  produced in RHIC and 
LHC are 
much smaller in comparison with  the  field strengths prevailed during the cosmological electro-weak phase transition which   may reach up to $eB\approx200m_\pi^2$
\cite{plb265}, 
they are strong enough to cast significant influence on the hadronic properties which bear the information of the chiral phase transition. 
 At vanishing chemical potential, modification due to the presence of magnetic background
can be obtained from first principle using lattice QCD simulations \cite{prd82M,prd83} which shows monotonic increase in critical
temperature with the increasing magnetic field. The effects of external magnetic field on the chiral phase transition 
has been studied using different effective models in recent years 
\cite{prd81,prd84,prc83,prd83M,prc83A,prd82R,prd83K,prd85,jhep08,prd82A,prd85S,prd84D,prd86E}.
QCD being a confining theory at low energies, effective theories are employed to describe the low energy behaviour of the strong interaction. 
In such a theory, the condensate is described as the non-zero expectation value of the sigma field which is basically a composite 
operator of two quark fields. If the condensate is already present without any background field, the effect of its  enhancement  in presence 
of the external magnetic field is described as \textit{magnetic catalysis}(MC).
Effective field theoretic models in general contain a few parameters which can be  fixed from experimental inputs. Although most of the model calculations are 
in support of 
MC,  some lattice results had shown inverse magnetic catalysis(IMC) where critical temperature follows
the opposite trend \cite{jhep1202,prdB86,jhep04,prd90}. It was  pointed out in \cite{jhep1304} that IMC is  attributed to the dominance of the sea 
contribution over the valence  contribution of the quark condensate.  
The sea  effect has not been incorporated even in the  Polyakov loop extended versions of  Nambu--Jona-Lasinio (PNJL) model 
and Quark-Meson(PQM) model which might be a possible reason for  the disagreement. To investigate the apparent contradiction, a significant amount of  work has
been done \cite{rmp88} in quest of proper modifications of the effective models, most of which are
focused on the magnetic field dependence of the coupling constants or other
 magnetic field dependent parameters in the model.  Very recently, 
IMC has been observed in   NJL model,
with Pauli-Villars regularization scheme \cite{plb758} which gives markedly different behaviour in comparison with the usual  soft-cutoff approach.


In the context of  nuclear physics, the MC effect was discussed  by Haber et al in Ref.\cite{Haber:2014ula}.
There, the effect of background magnetic field on 
the transition between vacuum to nuclear matter at zero temperature  was studied  for the Walecka model~\cite{walecka1}  as well as for the extended linear  sigma model. The study 
includes the B-dependent Dirac sea contribution of the free energy density  which was ignored previously (see for example
\cite{haber32,haber33,haber34,haber35,haber36,haber37,haber38,haber39,haber40}) in the case of magnetized nuclear matter. 
Following the renormalization  procedure similar to the case of magnetized quark matter, the cut-off dependence of the B-dependent sea contribution is absorbed 
into a  renormalized magnetic field and a renormalized electric charge. The onset of the vacuum to nuclear matter phase transition is determined by 
equating the corresponding free energies.  From the qualitative agreement between the two models, it is evident that with the  proper incorporation of 
the magnetic catalysis effect, the creation of the nuclear matter becomes energetically  more expensive in presence of the background magnetic field.
However, there exist an important qualitative difference between the two models. As the analysis suggests, only in case of the Walecka model, there exists a region where 
the critical chemical potential for the vacuum to nuclear matter transition  is lower than the same in the absence of the background field. This feature has surprising
similarities with the \textit{inverse magnetic catalysis}(IMC) shown in NJL and holographic Sakai-Sugimoto model \cite{haber15}. It is interesting to see whether 
similar feature exists also in a more generalized scenario. Now, as the anomalous  magnetic moment(AMM) of the nucleons
has not been taken into account in the analysis, an obvious generalization  will be to incorporate it in  the study of vacuum to nuclear matter phase 
transition under external magnetic field at non-zero temperature. A recent study \cite{referee} incorporating the magnetic field dependent 
vacuum in presence of finite temperature and density, however,  shows that the AMM of charged fermions makes no significant contribution to the equation of state at 
any external field value. Thus, among others, 
it will be interesting to see whether MC persists in the presence of anomalous magnetic moment.  

In this work  we restrict ourselves only in the ``weak'' field regime of the external magnetic field and  use the Walecka model to describe the nucleon-nucleon 
interaction. In this model, the interaction between the nucleons are 
described by the exchange of scalar ($\sigma$) and vector($\omega$) mesons. More realistic extension of the Walecka model where the  self-interactions of the meson fields
are also considered is ignored here for the sake of simplicity as they hardly contribute to the qualitative nature of the results presented in this work. Now, to obtain the 
effective mass of the nucleons, instead of minimizing the free energy density with respect to the condensate~\cite{Haber:2014ula}, 
we  calculate the effective nucleon propagator  by summing up the scalar and vector tadpole diagrams self-consistently. In that case, the effective mass 
of the nucleon appears as a pole of the effective nucleon propagator. In case of weak magnetic 
field, the nucleon propagators can be expressed as a series in powers of $qB$ and $\kappa B$ where $q$ and $\kappa$ represents the charge and the anomalous magnetic moment 
of the nucleons. It should be mentioned here that in the calculation of the tadpole diagrams using the interacting propagator, we employ mean 
field approximation. It 
 is  essentially equivalent to solving the meson field equations with the replacement of the meson field operators by their expectation values. In other words, under this 
 approximation, the meson field operators are rendered into classical fields assumed to be  uniform in space and time and the fluctuation around this background is neglected.  
 
The article is organized as follows. In Sec. \ref{sec.scalar.propagator}, the familiar expression \cite{ayala_scalar} of  the weak field expansion of the charged  
scalar propagator in presence of the constant  external magnetic field is derived using the perturbative method.  The same procedure is employed  in 
Sec. \ref{sec.fermion.propagator} to obtain the weak field expanded propagators  of the  charged and neutral fermion with non-zero magnetic moment. 
The suitable form of the corresponding thermal propagators are also discussed which are used to obtain  the effective mass of the 
nucleons in case of Walecka model described in Sec. \ref{sec.walecka}. Sec. \ref{num.reslt} contains the numerical results and discussions. Finally, we summarize 
our work in Sec. \ref{summary}.


\section{charged scalar propagator under external magnetic field} \label{sec.scalar.propagator}
Let us first consider the propagation of a charged scalar particle under \textit{zero external magnetic field}. 
In this case, the scalar vacuum Feynman propagator $\Delta_F\FB{x,x'}=\Delta_F\FB{x-x'}$ satisfies
\begin{eqnarray}
\FB{\partial_\mu\partial^\mu+m^2}\Delta_F\FB{x-x'}=\delta^4\FB{x-x'}~. \label{eq.KG.Green.1}
\end{eqnarray}
In order to solve Eq.~(\ref{eq.KG.Green.1}), we introduce the Fourier transform of $\Delta_F\FB{x-x'}$ by
\begin{eqnarray}
\Delta_F\FB{x-x'} = \int\frac{d^4k}{\FB{2\pi}^4}e^{-ik\cdot\FB{x-x'}}\Delta_F\FB{k}~, \label{eq.scalar.propagator.1}
\end{eqnarray}
where, $\Delta_F\FB{k}$ is the momentum space vacuum scalar propagator. 
Substituting $\Delta_F\FB{x-x'}$ from Eq.~(\ref{eq.scalar.propagator.1}) into Eq.~(\ref{eq.KG.Green.1}), we get
\begin{eqnarray}
\Delta_F\FB{k} = \FB{\frac{-1}{k^2-m^2+i\epsilon}}~,
\end{eqnarray}
where we have imposed the Feynman's boundary condition and put the $i\epsilon$ in the denominator.

We now turn on the \textit{external magnetic field}. In this case, the charged scalar propagator under external magnetic field, 
denoted by $\Delta_B\FB{x,x'}$ will satisfy,
\begin{eqnarray}
\TB{\SB{\frac{}{}\partial_\mu+iqA_\mu\FB{x}}\SB{\frac{}{}\partial^\mu+iqA^\mu\FB{x}}+m^2}\Delta_B\FB{x,x'} = \delta^4\FB{x-x'}~,
\label{eq.KG.Green.2}
\end{eqnarray}
where, $q$ is the electric charge of the particle and $A^\mu\FB{x}$ is the four potential corresponding 
to the external magnetic field. It is to be noted that, the propagator $\Delta_B\FB{x,x'}$ is not translational invariant. 
For solving Eq.~(\ref{eq.KG.Green.2}), we follow the procedure as given in Ref~\cite{Nieves:2004qp,Nieves:2006xp} and choose a particular gauge in which the four potential is 
\begin{eqnarray}
A^\mu\FB{x}=-\frac{1}{2}F^\munu x_\nu~. \label{eq.gauge.choice}
\end{eqnarray} 
For the case of a constant external magnetic field, the field strength tensor $F^\munu$ is independent of $x$. 
Substituting Eq.~(\ref{eq.gauge.choice}) into Eq.~(\ref{eq.KG.Green.2}), we get
\begin{eqnarray}
\TB{\Box+m^2-iqF^\munu x_\nu\partial_\mu - \frac{q^2}{4}F^{\mu\alpha}F_{\mu\beta}x_\alpha x^\beta}\Delta_B\FB{x,x'} = \delta^4\FB{x-x'}~. \label{eq.KG.Green.3}
\end{eqnarray}
The corresponding momentum space propagator $\Delta_B\FB{k}$ is obtained from the Fourier transform of the 
translational invariant part of the coordinate space propagator $\Delta_B\FB{x,x'}$ i.e.
\begin{eqnarray}
\Delta_B\FB{x,x'}=\phi\FB{x,x'}\int\frac{d^4k}{\FB{2\pi}^4}e^{-ik\cdot\FB{x-x'}}\Delta_B\FB{k}~, \label{eq.scalar.propagator.3}
\end{eqnarray}  
where, $\phi\FB{x,x'}$ is the phase factor and it depends on the choice of gauge. For the gauge given in 
Eq.~(\ref{eq.gauge.choice}), the phase factor comes out to be~\cite{Nieves:2006xp},
\begin{eqnarray}
\phi\FB{x,x'} = \exp\TB{\frac{i}{2}qF^\alphabeta x_\alpha x'^\beta}~.\label{eq.phase.factor}
\end{eqnarray}
Substituting Eq.~(\ref{eq.scalar.propagator.3}) into Eq.~(\ref{eq.KG.Green.3}), we get
\begin{eqnarray}
\int\frac{d^4k}{\FB{2\pi}^4}e^{-ik\cdot\FB{x-x'}}\Delta_B\FB{k}\TB{\Box+m^2-2ik_\mu\partial^\mu-k^2-qF^\munu x_\nu k_\mu-\frac{1}{4}q^2F^{\mu\alpha}F_{\mu\beta}x_\alpha x^\beta}\phi\FB{x,x'}=\delta^4\FB{x-x'}~. \label{eq.KG.Green.4}
\end{eqnarray}
We further substitute $\phi\FB{x,x'}$ from Eq.~(\ref{eq.phase.factor}) into Eq.~(\ref{eq.KG.Green.4}) and obtain
\begin{eqnarray}
\int\frac{d^4k}{\FB{2\pi}^4}e^{-ik\cdot\FB{x-x'}}\Delta_B\FB{k}\TB{-k^2+m^2-qk^\mu F_\munu\FB{x-x'}^\nu-\frac{1}{4}q^2F^{\mu\alpha}F_{\mu\beta}\FB{x-x'}_\alpha\FB{x-x'}^\beta}=\delta^4\FB{x-x'}~,
\label{eq.KG.Green.5}
\end{eqnarray}
where we have used the fact that $\phi\FB{x,x}=1$. This can be verified from Eq.~(\ref{eq.phase.factor}) 
using the antisymmetric property of $F^\munu$.
Each term in Eq.~(\ref{eq.KG.Green.5}) is now translationally invariant and can be expressed as,
\begin{eqnarray}
\int\frac{d^4k}{\FB{2\pi}^4}\Delta_B\FB{k}\TB{-k^2+m^2-iqk^\mu F_\munu\deltilde^\nu+\frac{1}{4}q^2F^{\mu\alpha}F_{\mu\beta}\deltilde_\alpha\deltilde^\beta}e^{-ik\cdot\FB{x-x'}}=\delta^4\FB{x-x'}~,
\label{eq.KG.Green.6}
\end{eqnarray} 
where we have used the notation $\deltilde_\mu = \FB{\frac{\del}{\del k^\mu}}$. 
In order to extract $\Delta_B\FB{k}$ from Eq.~(\ref{eq.KG.Green.6}), it is necessary to swap the positions of $\Delta_B\FB{k}$ 
and $e^{-ik\cdot\FB{x-x'}}$. This swapping is done at the cost of addition of a term, which 
contains a total four momentum derivative ($\deltilde_\mu J^\mu$) i.e.
\begin{eqnarray}
\int\frac{d^4k}{\FB{2\pi}^4}e^{-ik\cdot\FB{x-x'}}\TB{-k^2+m^2+iq F_\munu\deltilde^\nu k^\mu+\frac{1}{4}q^2F^{\mu\alpha}F_{\mu\beta}\deltilde_\alpha\deltilde^\beta}\Delta_B\FB{k} + \int\frac{d^4k}{\FB{2\pi}^4}\deltilde_\mu J^\mu=\delta^4\FB{x-x'}~.
\end{eqnarray}
$J^\mu$ in the above equation contains the propagator $\Delta_B\FB{k}$. 
Now the $d^4k$ integral of second term on the L.H.S. can be converted to a surface integral using Gauss's theorem 
and assuming $D\FB{k}$ to be well behaved function, this term will vanish. So the momentum space propagator $\Delta_B\FB{k}$ 
satisfies the following differential equation,
\begin{eqnarray}
\TB{-k^2+m^2+iq F_\munu\deltilde^\nu k^\mu+\frac{1}{4}q^2F^{\mu\alpha}F_{\mu\beta}\deltilde_\alpha\deltilde^\beta}\Delta_B\FB{k} = 1
\label{eq.scalar.propagator.4}
\end{eqnarray}
Let us now consider a constant external magnetic field in the +ve z-direction i.e. $\vec{B}=B\hat{z}$, which implies that 
the non-zero components of the electromagnetic field strength tensor $F^\munu$ are $F^{12}=-F^{21}$. 
So any four vector $a^\mu\equiv\FB{a^0,a^1,a^2,a^3}$ is decomposed into 
$a^\mu=\FB{a_\parallel^\mu+a_\perp^\mu}$ with $a^\mu_\parallel\equiv\FB{a^0,0,0,a^3}$ and 
$a^\mu_\perp\equiv\FB{0,a^1,a^2,0}$. The corresponding metric tensors are $g_\parallel^\munu=diag\FB{1,0,0,-1}$ and 
$g_\perp^\munu=diag\FB{0,-1,-1,0}$ satisfying $g^\munu=\FB{g_\parallel^\munu+g_\perp^\munu}$.  
Therefore the propagator $\Delta_B\FB{k}$ being a Lorentz scalar, will be functions of $k_\parallel^2$ and $k_\perp^2$ i.e. 
$\Delta_B\FB{k}=\Delta_B\FB{k_\parallel^2,k_\perp^2}$. Hence the third term within the square bracket in the L.H.S. of 
Eq.~(\ref{eq.scalar.propagator.4}) can be written as,
\begin{eqnarray}
iqF_\munu\deltilde^\nu\TB{k^\mu\Delta_B\FB{k_\parallel^2,k_\perp^2}} 
= iqF_\munu\TB{g^\munu+2k^\mu k_\parallel^\nu\frac{\del}{\del k_\parallel^2}
	+2k^\mu k_\perp^\nu\frac{\del}{\del k_\perp^2}}\Delta_B\FB{k_\parallel^2,k_\perp^2} =0~.
\end{eqnarray} 
It is also trivial to check that
\begin{eqnarray}
F^{\mu\alpha}F_{\mu\beta}\deltilde_\alpha\deltilde^\beta = B^2g_\perp^\alphabeta\deltilde_\alpha\deltilde_\beta 
= B^2\tilde{\Box}_\perp~,
\end{eqnarray}
where, $\tilde{\Box}_\perp = g_\perp^\munu\deltilde_\alpha\deltilde_\beta$. Finally Eq.~(\ref{eq.scalar.propagator.4}) becomes,
\begin{eqnarray}
\TB{-k^2+m^2+\frac{1}{4}\FB{qB}^2\tilde{\Box}_\perp}\Delta_B\FB{k} = 1~. \label{eq.scalar.propagator.5}
\end{eqnarray}
We now expand the propagator as a power series in $qB$, 
\begin{eqnarray}
\Delta_B\FB{k} = \sum_{i=0}^{\infty}\FB{qB}^i\Delta_i\FB{k}
\end{eqnarray}
and substitute in Eq.~(\ref{eq.scalar.propagator.5}) to obtain,
\begin{eqnarray}
\sum_{i=0}^{\infty}\TB{\FB{qB}^i\FB{-k^2+m^2}+\FB{qB}^{i+2}\frac{1}{4}\tilde{\Box}_\perp}\Delta_i\FB{k}=1~.
\end{eqnarray}
Equating the coefficients of different powers of $qB$ in the both side of the above equation, we get,
\begin{eqnarray}
\Delta_0\FB{k}&=&\FB{\frac{-1}{k^2-m^2+i\epsilon}} \nn \\
\Delta_1\FB{k}&=& 0 \nn \\
\Delta_n\FB{k}&=& -\Delta_0\FB{k}\frac{1}{4}\tilde{\Box}_\perp\Delta_{n-2}\FB{k} ~~\text{for}~~ n\geq2~. \label{eq.scalar.propagator.recursion}
\end{eqnarray} 
Eq.~(\ref{eq.scalar.propagator.recursion}) is a recursion relation and it immediately follows that 
$\Delta_3\FB{k}=\Delta_5\FB{k}=\Delta_7\FB{k} =.....= 0$. Using this relation one can calculate the charged scalar propagator 
up to any order in $eB$. As for example,
\begin{eqnarray}
\Delta_2\FB{k}=-\Delta_0\FB{k}\frac{1}{4}\tilde{\Box}_\perp\Delta_0\FB{k} 
=\TB{\frac{k_\parallel^2-k_\perp^2-m^2}{\FB{k^2-m^2+i\epsilon}^4}}~. \nn
\end{eqnarray}
Therefore the propagator becomes,
\begin{eqnarray}
\Delta_B\FB{k}&=&\Delta_0\FB{k}+\FB{qB}^2\Delta_2\FB{k}+\mathcal{O}\FB{qB}^4 \nn \\
&=& \FB{\frac{-1}{k^2-m^2+i\epsilon}} + \FB{qB}^2\TB{\frac{k_\parallel^2-k_\perp^2-m^2}{\FB{k^2-m^2+i\epsilon}^4}}
+\mathcal{O}\FB{qB}^4~.
\end{eqnarray}


\section{fermion propagator under external magnetic field}
\label{sec.fermion.propagator}

Following the similar procedure described in the previous section, we find that the Dirac equation with anomalous magnetic moment ($\kappa$)
in the momentum space representation is given by
\cite{Nieves:2004qp,Nieves:2006xp}  
\bea
\Big[\ps -\frac{i}{2}q F^{\mu\nu}\g_\mu\frac{\p}{\p p^\nu}-\m-\frac{1}{2}\k \s\cdot F\Big]S_B(p)=1.
\eea

The strategy to obtain the power expansion is to write 
\bea
S_B&=&S_0+S_1.
\eea
where  $S_0$ represents the vacuum propagator and $S_1$ represents its linear order correction in presence of external magnetic field. Now, 
let us define the operator 
\bea
\op&=&\Big[\frac{i}{2}q F^{\mu\nu}\g_\mu\frac{\p}{\p p^\nu}+\frac{1}{2}\k \s\cdot F\Big]
\eea
 Using the perturbative expansion in the Dirac equation and neglecting the higher order $\op S_1$ term one obtains  
\bea
S_1&=&S_0\op S_0.
\eea
Thus the linear order correction to the weak expansion of the propagator is nothing but an operator of non-commutative gamma matrices and differentials
sandwiched between the familiar vacuum propagators. 
Following the similar strategy one can extend the series to  higher order terms in powers of $B$. As we shall see that in our case, 
the  leading order contribution of the external magnetic field occurs due to the  quadratic correction of the weak field propagator 
and not due to  the simpler linear order one, 
we must extend the perturbative series as 
\bea
S_B&=&S_0+S_1+S_2
\eea
for which one  obtains 
\bea
S_2&=&S_0\op S_1
\eea
where  $S_1=S_0 \op S_0$  is  given by ( see \cite{Nieves:2004qp,Nieves:2006xp} )
\bea
S_1&=&
\frac{1}{(p^2-\m^2+i\e)^2}\nn\\
&\times&\Big[qB\gf\big[(p\cdot b)\us-(p\cdot u)\bs
+\m\us\bs\big]+\k B\big[(\ps + \m)\gf\us\bs(\ps + \m)\big]\Big]
\eea
with $u^\mu\equiv(1,0,0,0)$ and $b^\mu\equiv(0,0,0,1)$ in the fluid rest frame. It is straightforward  to derive  the expression of 
$S_2$ and after plugging the correction terms we finally obtain   the weak field expansion of the  fermion propagator given by  

\begin{eqnarray}
S_B\FB{p,m}&=&\frac{-\FB{\cancel{p}+m}}{p^2-m^2+i\epsilon} 
+ \FB{qB}\frac{i\gamma^1\gamma^2\FB{\cancel{p}_\parallel+m}}{\FB{p^2-m^2+i\epsilon}^2}
+\FB{\kappa B}\frac{\FB{\cancel{p}+m}i\gamma^1\gamma^2\FB{\cancel{p}+m}}{\FB{p^2-m^2+i\epsilon}^2} \nn \\
&& +\FB{qB}^2\frac{-2\SB{p_\perp^2\FB{\cancel{p}_\parallel+m}-\cancel{p}_\perp\FB{p_\parallel^2-m^2}}}{\FB{p^2-m^2+i\epsilon}^4}
+ \FB{qB}\FB{\kappa B}\frac{-4\cancel{p}_\parallel\FB{\cancel{p}_\parallel+m}+p^2-m^2}{\FB{p^2-m^2+i\epsilon}^3} \nn \\
&& + \FB{\kappa B}^2\frac{-\FB{\cancel{p}+m}\FB{\cancel{p}_\parallel-\cancel{p}_\perp+m}\FB{\cancel{p}+m}}{\FB{p^2-m^2+i\epsilon}^3}
+ \mathcal{O}\FB{B^3}~. \label{eq.fermion.propagator.10}
\end{eqnarray}
In order to express $S_B\FB{p,m}$ in a more compact form, we use the procedure given in Ref.~\cite{Bandyopadhyay:2017raf} and 
write
\begin{eqnarray}
\FB{\frac{-1}{p^2-m^2+i\epsilon}}^n = \left.\hat{A}_{n-1}\Delta_F\FB{p,m_1}\right|_{m_1=m}
\label{eq.fer.nn.1}
\end{eqnarray}
where,
\begin{eqnarray}
\Delta_F\FB{p,m}=\FB{\frac{-1}{p^2-m^2+i\epsilon}} 
\end{eqnarray}
and 
\begin{eqnarray}
\hat{A}_n = \frac{\FB{-1}^n}{n!}\frac{\del^n}{\del\FB{m_1^2}^n}~. \label{eq.A.hat}
\end{eqnarray}
Using Eqs.~(\ref{eq.fer.nn.1})-(\ref{eq.A.hat}), we can rewrite Eq.~(\ref{eq.fermion.propagator.10}) as
\begin{eqnarray}
S_B\FB{p,m} = \left.\hat{F}\FB{p,m,m_1}\Delta_F\FB{p,m_1}\right|_{m_1=m} \label{eq.fermion.propagator.11}
\end{eqnarray}
where,
\begin{eqnarray}
\hat{F}\FB{p,m,m_1}&=&\FB{\cancel{p}+m} 
+ \FB{qB}i\gamma^1\gamma^2\FB{\cancel{p}_\parallel+m}\hat{A}_1
+\FB{\kappa B}\FB{\cancel{p}+m}i\gamma^1\gamma^2\FB{\cancel{p}+m}\hat{A}_1 \nn \\ &&
-2\FB{qB}^2\SB{p_\perp^2\FB{\cancel{p}_\parallel+m}-\cancel{p}_\perp\FB{p_\parallel^2-m^2}}\hat{A}_3
+ \FB{qB}\FB{\kappa B}\SB{4\cancel{p}_\parallel\FB{\cancel{p}_\parallel+m}-p^2+m^2}\hat{A}_2 \nn \\ &&
+ \FB{\kappa B}^2\FB{\cancel{p}+m}\FB{\cancel{p}_\parallel-\cancel{p}_\perp+m}\FB{\cancel{p}+m}\hat{A}_2
+ \mathcal{O}\FB{B^3}~. \label{eq.F}
\end{eqnarray}

We conclude this section by mentioning the fermion propagator at finite temperature and density along with 
external magnetic field. For this, we use the real time formalism  of thermal field theory where the thermal propagator becomes $2\times2$ matrix. 
However, it is sufficient to know the $11$-component of the matrix propagator \cite{Mallik:2016anp} which 
 is given by~\cite{DOlivo:2002omk},
\begin{eqnarray}
S_{11}\FB{p,m} = S_B\FB{p}-\eta\FB{p\cdot u}\TB{S_B\FB{p}-\gamma^0S_B^\dagger\FB{p}\gamma^0} \label{eq.S11.1}
\end{eqnarray}
where,
\begin{eqnarray}
\eta\FB{p\cdot u}=\Theta\FB{p\cdot u}f_+\FB{p\cdot u} + \Theta\FB{-p\cdot u}f_-\FB{-p\cdot u} \label{eq.eta}
\end{eqnarray}
with
\begin{eqnarray}
f_\pm\FB{p\cdot u} = \TB{\exp\FB{\frac{p\cdot u\mp \mu}{T}}+1}^{-1}~. \label{eq.f.pm}
\end{eqnarray}
Substituting Eq.~(\ref{eq.fermion.propagator.11}) into Eq.~(\ref{eq.S11.1}) and using the fact that 
$\gamma^0\hat{F}^\dagger\FB{p,m,m_1}\gamma^0=\hat{F}\FB{p,m,m_1}$, we get
\begin{eqnarray}
S_{11}\FB{p,m} = \left.\hat{F}\FB{p,m,m_1}\TB{\frac{}{}\Delta_F\FB{p,m_1}-2\pi i\eta\FB{p\cdot u}\delta\FB{p^2-m_1^2}}\right|_{m_1=m}
\label{eq.S11.3}
\end{eqnarray}


\section{Effective mass of nucleon in Walecka model} \label{sec.walecka}

The propagation of nucleons in hot and dense nuclear matter is well described using Quantum Hadrodynamics (QHD) details of 
which can be found in Ref.~\cite{Negele:1986bp, Alam:1999sc}. We briefly summarize the main the formalism of QHD at \textit{zero magnetic field}. 
We start with the real time thermal propagator matrix of the nucleon ~\cite{Mallik:2016anp,bellac},
\begin{eqnarray}
\bm{S}_0\FB{p,m_N}=\FB{\cancel{p}+m_N} \bm{V}\left[\begin{array}{cc}
\Delta_F\FB{p,m_N} & 0 \\
0 & -\Delta_F^*\FB{p,m_N}
\end{array}\right]\bm{V}
\end{eqnarray}
where the diagonalizing matrix $\bm{V}$ is given by,
\begin{eqnarray}
\bm{V} = \left[\begin{array}{cc}
N_2 & -N_1e^{\beta\mu/2} \\
N_1e^{-\beta\mu/2} & N_2
\end{array}\right]
\end{eqnarray}
with 
\begin{eqnarray}
N_1\FB{p\cdot u} &=& \sqrt{f_+\FB{p\cdot u}}\Theta\FB{p\cdot u} + \sqrt{f_-\FB{-p\cdot u}}\Theta\FB{-p\cdot u} \nn \\
N_2\FB{p\cdot u} &=& \sqrt{1-f_+\FB{p\cdot u}}\Theta\FB{p\cdot u} + \sqrt{1-f_-\FB{-p\cdot u}}\Theta\FB{-p\cdot u}~. \nn
\end{eqnarray}

In Walecka model, the nucleons interact with the scalar meson $\sigma$ and vector meson $\omega$. The interaction Lagrangian is 
\begin{eqnarray}
\scrL_{QHD} = g_{\sigma NN}\bar{\Psi}\Psi\sigma -g_{\omega NN}\bar{\Psi}\gamma^\mu\Psi\omega_\mu~, \label{eq.lagrangian.walecka}
\end{eqnarray}
where $\Psi=\left[\begin{array}{c} p \\ n \end{array} \right]$ is the nucleon isospin doublet and the value 
of the coupling constants are given by $g_{\sigma NN}=9.57$ and $g_{\omega NN}=11.67$~\cite{Negele:1986bp}. 
The complete nucleon propagator matrix $\bm{S'}\FB{p,m_N}$ in presence of these interactions is obtained from 
the Dyson-Schwinger equation given by,
\begin{eqnarray}
\bm{S'} = \bm{S}_0 - \bm{S}_0\bm{\Sigma}\bm{S'} \label{eq.dyson}
\end{eqnarray}
where, $\bm{\Sigma}$ is the one-loop thermal self energy matrix of the nucleon. It can be shown that~\cite{Mallik:2016anp}, 
the complete propagator and the self energy matrices are diagonalized by $\bm{V}$ and $\bm{V}^{-1}$ respectively. This in turn 
diagonalizes the Dyson-Schwinger equation and Eq.~(\ref{eq.dyson}) becomes an algebraic equation (in thermal space),
\begin{eqnarray}
\overline{S'} = \overline{S}_0 - \overline{S}_0\overline{\Sigma}~\overline{S'}~. \label{eq.dyson.diagonal}
\end{eqnarray}
It is to be noted that, each term in the above equation is $4\times4$ matrix in Dirac space. Here 
$\overline{S}_0\FB{p,m_N}=-\FB{\cancel{p}+m_N}\Delta_F\FB{p,m_N}$ and $\overline{\Sigma}$ is 
the $11$-component of the matrix $\bm{V}^{-1}\bm{\Sigma}\bm{V}^{-1}$ and is called the thermal self energy function. 
In Walecka model the Dirac structure of $\overline{\Sigma}$ comes out to be,
\begin{eqnarray}
\overline{\Sigma} = \FB{\Sigma_s\mathds{1}+\Sigma_v^\mu\gamma_\mu} = \FB{\Sigma_s\mathds{1}+\cancel{\Sigma}_v}~.
\label{eq.dirac.struct}
\end{eqnarray}
Using Eq.~(\ref{eq.dirac.struct}), we can solve Eq.~(\ref{eq.dyson.diagonal}) and obtain
\begin{eqnarray}
\overline{S'}\FB{p,m_N} = \FB{\cancel{P}+m_N^*}\Delta_F\FB{P,m_N^*} 
\end{eqnarray}
where
\begin{eqnarray}
P=\FB{p-\Sigma_v} ~~~~ \text{and} ~~~~~ m_N^* = \FB{m+\Sigma_s}~.
\end{eqnarray}
We can finally write down the complete propagator matrix 
\begin{eqnarray}
\bm{S'}\FB{p,m_N}=\FB{\cancel{P}+m_N^*} \bm{V}\left[\begin{array}{cc}
\Delta_F\FB{P,m_N^*} & 0 \\
0 & -\Delta_F^*\FB{P,m_N^*}
\end{array}\right]\bm{V}~,
\label{eq.S'}
\end{eqnarray}
whose $11$-component is,
\begin{eqnarray}
S'_{11}\FB{p,m_N} = S_F\FB{P,m_N^*}-\eta\FB{P\cdot u}\TB{S_F\FB{P,m_N^*}-\gamma^0S_F^\dagger\FB{P,m_N^*}\gamma^0}~. \label{eq.S11.2}
\end{eqnarray}

\begin{figure}[h]
\includegraphics[scale=0.8]{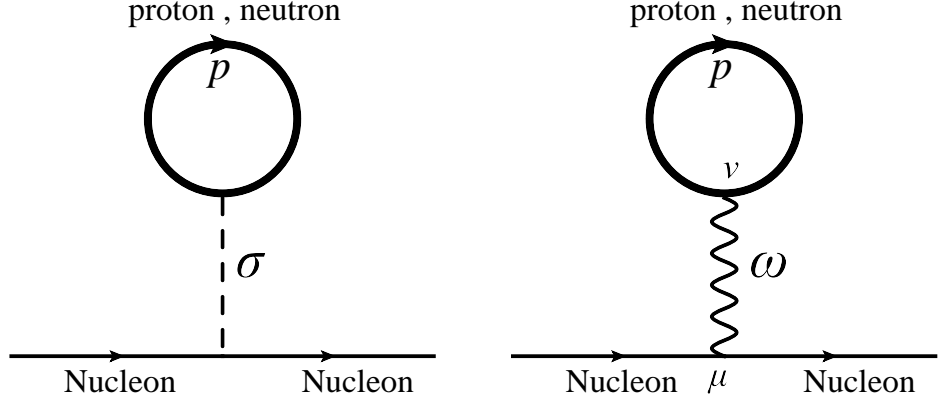}
\caption{Feynman diagrams for the one-loop self energy of nucleon in Walecka model. Bold line indicates the 
	complete/dressed propagator}
\label{fig.walecka.feynman.diagram}
\end{figure}

Let us now calculate, the nucleon self energy function $\bar{\Sigma}$ using the interaction Lagrangian given in 
Eq.~(\ref{eq.lagrangian.walecka}) and consider only the tadpole Feynman diagrams as shown in 
Fig.~\ref{fig.walecka.feynman.diagram}. It is to be 
noted that, the loop particles are dressed i.e. the propagator for the loop particles is $\bm{S'}\FB{p,m_N}$ as given in 
Eq.~(\ref{eq.S'}). Applying Feynman rule to Fig.~\ref{fig.walecka.feynman.diagram} we obtain the $11$-component of the thermal self energy as,
\begin{eqnarray}
\Sigma_{11} &=& -\FB{\frac{g^2_{\sigma NN}}{m_\sigma^2}}i\int\frac{d^4p}{\FB{2\pi}^4}
\Tr\TB{S'^{~(\text{p})}_{11}\FB{p,m_N}+S'^{~(\text{n})}_{11}\FB{p,m_N}} \nn \\
&& + \gamma_\mu \FB{\frac{g^2_{\omega NN}}{m_\omega^2}}i\int\frac{d^4p}{\FB{2\pi}^4}
\Tr\TB{\gamma^\mu S'^{~(\text{p})}_{11}\FB{p,m_N}+ \gamma^\mu S'^{~(\text{n})}_{11}\FB{p,m_N}}
\end{eqnarray}
where (p) and (n) in the superscript corresponds to proton and neutron respectively.
It is easy to show that $\RE\Sigma_{11}=\RE\overline{\Sigma}$. So we get from Eq.~(\ref{eq.dirac.struct}),
\begin{eqnarray}
\RE\Sigma_s &=& -\FB{\frac{g^2_{\sigma NN}}{m_\sigma^2}}\RE ~i\int\frac{d^4p}{\FB{2\pi}^4}
\Tr\TB{S'^{~(\text{p})}_{11}\FB{p,m_N}+S'^{~(\text{n})}_{11}\FB{p,m_N}} \label{eq.sigma.s} \\
\RE\Sigma_v^\mu &=& \FB{\frac{g^2_{\omega NN}}{m_\omega^2}}\RE ~i\int\frac{d^4p}{\FB{2\pi}^4}
\Tr\TB{\gamma^\mu S'^{~(\text{p})}_{11}\FB{p,m_N}+ \gamma^\mu S'^{~(\text{n})}_{11}\FB{p,m_N}}~. \label{eq.sigma.v}
\end{eqnarray}
Substituting $S'_{11}\FB{p,m_N}$ from Eq.~(\ref{eq.S11.2}) into Eqs.~(\ref{eq.sigma.s}) and (\ref{eq.sigma.v}) 
and performing the $dp^0$ integral, we get
\begin{eqnarray}
\RE\Sigma_s\FB{m_N^*} &=& \RE\Sigma_s^{(\text{pure vacuum})}-\FB{\frac{4g^2_{\sigma NN}m_N^*}{m_\sigma^2}}\int\frac{d^3p}{\FB{2\pi}^3}\FB{\frac{1}{\Omega_p}}
\TB{N^p_+ + N^p_-} \label{eq.sigma.s.2}\\ 
\RE\Sigma_v^\mu\FB{m_N^*} &=& \FB{\frac{4g^2_{\omega NN}}{m_\omega^2}}\int\frac{d^3p}{\FB{2\pi}^3}
\TB{N^p_+ - N^p_-}\delta_0^\mu
\end{eqnarray}
where, $\Omega_p = \sqrt{\vec{p}^2+\FB{m_N^*}^2}$ and 
\begin{eqnarray}
N^p_\pm = \TB{\exp\FB{\frac{\Omega_p\mp \mu}{T}}+1}^{-1}~.
\end{eqnarray}
In Eq.~(\ref{eq.sigma.s.2}), $\RE\Sigma_s^{(\text{pure vacuum})}$ is given by
\begin{eqnarray}
\RE\Sigma_s^{(\text{pure vacuum})} = \FB{\frac{8m_N^*g_{\sigma NN}^2}{m_\sigma^2}} \RE ~i\int\frac{d^4p}{\FB{2\pi}^4}
\TB{\frac{1}{p^2-\FB{m_N^*}^2+i\epsilon}}~. \label{eq.sigma.s.purevacuum}
\end{eqnarray}
We will neglect the contribution of vacuum self energy term $\RE\Sigma_s^{(\text{pure vacuum})}$ in 
Eq.~(\ref{eq.sigma.s.2}) following the  Mean Field Theory (MFT)~\cite{Negele:1986bp} approach.

The effective mass of the nucleon ($m_N^*$) can be calculated from the pole of the complete nucleon propagator which essentially 
means solving the self consistent equation,
\begin{eqnarray}
m_N^* = m_N+\RE\Sigma_s\FB{m_N^*}~. \label{eq.trans}
\end{eqnarray}
%

Let us now turn on the \textit{external magnetic field}. Since we are only interested in the effective mass of 
nucleon, let us calculate the scalar self energy $\RE\Sigma_s\FB{m_N^*}$. In this case, the proton and neutron 
propagators in Eq.~(\ref{eq.sigma.s}) have to be replaced as $S'_{11}\FB{p,m_N} \rightarrow S_{11}\FB{P,m_N^*}$
where $S_{11}\FB{p,m}$ is defined in Eq.~(\ref{eq.S11.3}). This implies, 
\begin{eqnarray}
S'^{~(\text{p,n})}_{11}\FB{p,m_N} &=& \left.\hat{F}^{(\text{p,n})}\FB{P,m_N^*,m_1}\TB{\frac{}{}\Delta_F\FB{P,m_1}-
2\pi i\eta\FB{P\cdot u}\delta\FB{P^2-m_1^2}}\right|_{m_1=m_N^*} \label{eq.S11'.pn} 
\end{eqnarray}
where $\hat{F}^{(\text{p})}\FB{p,m,m_1}$ and $\hat{F}^{(\text{n})}\FB{p,m,m_1}$ 
are obtained from Eq.~(\ref{eq.F}) by replacing 
$q$ and $\kappa$ with the corresponding values of proton and neutron respectively i.e. 
for proton $q\rightarrow\MB{e}, \kappa\rightarrow\kappa_{\text{p}}$ and for neutron $q\rightarrow0, \kappa\rightarrow\kappa_{\text{n}}$. Here $\MB{e}$ is the absolute electronic charge and the anomalous magnetic moments 
of proton and neutron are given by $\kappa_{\text{p}} = g_{\text{p}}\FB{\frac{\MB{e}}{2m_N}}$ 
and $\kappa_{\text{n}} = g_{\text{n}}\FB{\frac{\MB{e}}{2m_N}}$ respectively with 
$g_{\text{p}}=1.79, g_{\text{n}}=-1.91$.

Substituting Eq.~(\ref{eq.S11'.pn}) into Eq.~(\ref{eq.sigma.s}), and shifting the momentum 
$p\rightarrow \FB{p+\Sigma_V}$, we get
\begin{eqnarray}
\RE\Sigma_s &=& -\FB{\frac{g^2_{\sigma NN}}{m_\sigma^2}}\RE ~i\int\frac{d^4p}{\FB{2\pi}^4}
\Tr\left.\TB{\hat{F}^{(\text{p})}\FB{p,m_N^*,m_1}+\hat{F}^{(\text{n})}\FB{p,m_N^*,m_1}}\Delta_F\FB{p,m_1}\right|_{m_1=m_N^*}
\nn \\ && -\FB{\frac{g^2_{\sigma NN}}{m_\sigma^2}}\int\frac{d^4p}{\FB{2\pi}^4}
\left.\frac{}{}\Tr\TB{\hat{F}^{(\text{p})}\FB{p,m_N^*,m_1}+\hat{F}^{(\text{n})}\FB{p,m_N^*,m_1}} 2\pi\eta\FB{p\cdot u}\delta\FB{p^2-m_1^2}\right|_{m_1=m_N^*} \\ 
\implies \RE\Sigma_s &=& \Sigma_s^\text{(vacuum)} + \Sigma_s^\text{(medium)} \label{eq.sigma.s.full}
\end{eqnarray}
where,
\begin{eqnarray}
\Sigma_s^{\text{(vacuum)}} &=& -\FB{\frac{g^2_{\sigma NN}}{m_\sigma^2}}\RE ~i\int\frac{d^4p}{\FB{2\pi}^4}
\hat{T}\FB{p,m_N^*,m_1}\left.\frac{}{}\Delta_F\FB{p,m_1}\right|_{m_1=m_N^*} \label{eq.sigma.s.B} \\ 
\Sigma_s^\text{(medium)} &=& -\FB{\frac{g^2_{\sigma NN}}{m_\sigma^2}} \int\frac{d^4p}{\FB{2\pi}^4}
\hat{T}\FB{p,m_N^*,m_1}\left.\frac{}{}2\pi \eta\FB{p\cdot u}\delta\FB{p^2-m_1^2}\right|_{m_1=m_N^*} \label{eq.sigma.sp.1}~.
\end{eqnarray}
In the above equations,
\begin{eqnarray}
\hat{T}\FB{p,m_N^*,m_1} &=& \Tr\TB{\hat{F}^{(\text{p})}\FB{p,m_N^*,m_1}+\hat{F}^{(\text{n})}\FB{p,m_N^*,m_1}} \nn \\
&=& 8m_N^* - 8m_N^*\FB{eB}^2p_\perp^2\hat{A}_3 + 4m_N^*\SB{\FB{\kappa_{\text{p}}B}^2+\FB{\kappa_{\text{n}}B}^2}
\SB{\FB{m_N^*}^2+p^2-2p_\perp^2+2p_\parallel^2}\hat{A}_2 \nn \\ && \hspace{2cm}
+~ 4 \FB{\MB{e}B}\FB{\kappa_{\text{p}}B}\SB{\FB{m_N^*}^2-p^2+4p_\parallel^2}\hat{A}_2 
\label{eq.T}
\end{eqnarray}
The detailed calculation of $\Sigma_s^{\text{(vacuum)}}$ and $\Sigma_s^{\text{(medium)}}$ are provided in 
Appendices~\ref{appendix.sigma.s.vacuum} and \ref{appendix.sigma.s.BT}. The expression for $\Sigma_s^{\text{(vacuum)}}$  
can be read off Eq.~(\ref{eq.sigmavac.final}) as
\begin{eqnarray}
\Sigma_s^{\text{(vacuum)}} &=& \FB{\frac{g^2_{\sigma NN}}{4\pi^2m_\sigma^2}}
\TB{\frac{\FB{eB}^2}{3m_N^*}+\SB{\FB{\kappa_{\text{p}}B}^2m_N^*+\FB{\kappa_{\text{n}}B}^2m_N^* 
		+\FB{\MB{e}B}\FB{\kappa_{\text{p}}B}}\SB{\frac{1}{2}+2\ln\FB{\frac{m_N^*}{m_N} } }}~. \label{eq.sigmavac}
\end{eqnarray}
The calculation of $\Sigma_s^\text{(medium)}$ 
is performed for two different cases separately, namely (1) the zero temperature case and (2) the finite temperature case. 
For \textit{zero temperature}, we have from Eq.~(\ref{eq.sigma.spn.final})
\begin{eqnarray}
\Sigma_s^\text{(medium)} &=& -\FB{\frac{2g^2_{\sigma NN}}{\pi^2m_\sigma^2}} \TB{m_N^*I_2\FB{\mu_\text{B},m_N^*}+\frac{1}{3}\FB{eB}^2m_N^*
	C_1\FB{\mu_\text{B},m_N^*} \right. \nn \\ && \left.
	+ 2\SB{m_N^*\FB{\kappa_\text{p}B}^2+m_N^*\FB{\kappa_\text{n}B}^2+\FB{\MB{e}B}\FB{\kappa_\text{p}B}}\SB{m_N^{*2}C_1\FB{\mu_\text{B},m_N^*}
		+\frac{1}{3}C_2\FB{\mu_\text{B},m_N^*}} } \label{eq.sigma.spn} ~.
\end{eqnarray}
Whereas For \textit{finite temperature}, we have from (\ref{eq.sigma.spnt.final}),
\begin{eqnarray}
\Sigma_s^{\text{(medium)}} &=& -\FB{\frac{2g^2_{\sigma NN}}{\pi^2m_\sigma^2}} \int_{0}^{\infty}\MB{\vec{p}}^2d\MB{\vec{p}}
\TB{m_N^*\FB{\tilde{C}_1^{+p}+\tilde{C}_1^{-p}}+\frac{2}{3}m_N^*\FB{eB}^2\MB{\vec{p}}^2
	\FB{\tilde{C}_3^{+p}+\tilde{C}_3^{-p}} \right. \nn \\ 
	&& \left. +2\FB{m_N^{*2}+\frac{2}{3}\MB{\vec{p}}^2}\SB{m_N^*\FB{\kappa_\text{p}B}^2+m_N^*\FB{\kappa_\text{n}B}^2+\FB{\MB{e}B}
		\FB{\kappa_\text{p}B}}
	\FB{\tilde{C}_2^{+p}+\tilde{C}_2^{-p}}} \label{eq.sigma.spnt}
\end{eqnarray}
The definition of the functions $I_2$, $C_1$, $C_2$, $\tilde{C}^\pm_1$, $\tilde{C}^\pm_2$ and $\tilde{C}^\pm_3$ can be found 
in Appendix~\ref{appendix.sigma.s.BT}.


\section{Numerical Results}
\label{num.reslt}

\begin{figure}[h]
	\includegraphics[scale=0.3,angle=-90]{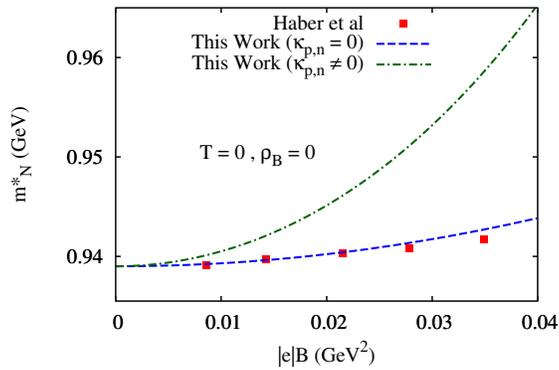}
	\caption{Variation of $m_N^*$ with $\MB{e}B$ at \textit{zero temperature and zero density}. Results with and without the 
		anomalous magnetic moment of nucleons are compared with results from Ref.\cite{Haber:2014ula}.}
	\label{fig.mstar.t0rho0}
\end{figure}

 We begin this section by obtaining  the  effective nucleon mass with external magnetic field at 
 \textit{zero temperature and zero density}. 
 In this case the contribution from $\Sigma_s^{\text{(medium)}}=0$. 
 Thus we need to solve the transcendental equation,
 \begin{eqnarray}
 m_N^* = m_N + \Sigma_s^{\text{(vacuum)}}\FB{m_N^*} \label{eq.trans.1}
 \end{eqnarray}
 where, $\Sigma_s^{\text{(vacuum)}}\FB{m_N^*}$ is given in Eq.~(\ref{eq.sigmavac}). At first  we  neglect the 
 effect of anomalous magnetic moment of nucleons so that the above equation simplifies to 
 \begin{eqnarray}
 m_N^* = m_N + \frac{g^2_{\sigma NN}\FB{eB}^2}{12\pi^2m_\sigma^2m_N^*}
 \end{eqnarray}
 which can be solved  analytically to obtain 
 \begin{eqnarray}
 m_N^* = \frac{1}{2}\TB{m_N+\sqrt{m_N^2+\frac{g^2_{\sigma NN}\FB{eB}^2}{3\pi^2m_\sigma^2}}}~.
 \end{eqnarray}
 As can be seen from the above equation, the effective nucleon mass increases monotonically with the 
 increase of $eB$. This enhancement is shown in 
 Fig.~\ref{fig.mstar.t0rho0} where it is also  compared with the result from Ref~\cite{Haber:2014ula}. Though the current approach to  
 obtain the effective nucleon mass differs  from Ref~\cite{Haber:2014ula}, there exists a noticeable  quantitative agreement between the two results 
 in the weak magnetic field regime. 
 Now we include the anomalous magnetic moments of nucleons 
 and solve Eq.~(\ref{eq.trans.1}) numerically. It is found that the incorporation of nucleon magnetic moment further increases the effective 
 mass   and this  effect remains significant even in case of weak magnetic fields as shown in
 Fig.~\ref{fig.mstar.t0rho0}. In other words, the nucleon magnetic moment favors the magnetic catalysis effect at zero temperature and zero baryon density.

\begin{figure}[h]
	\includegraphics[scale=0.3,angle=-90]{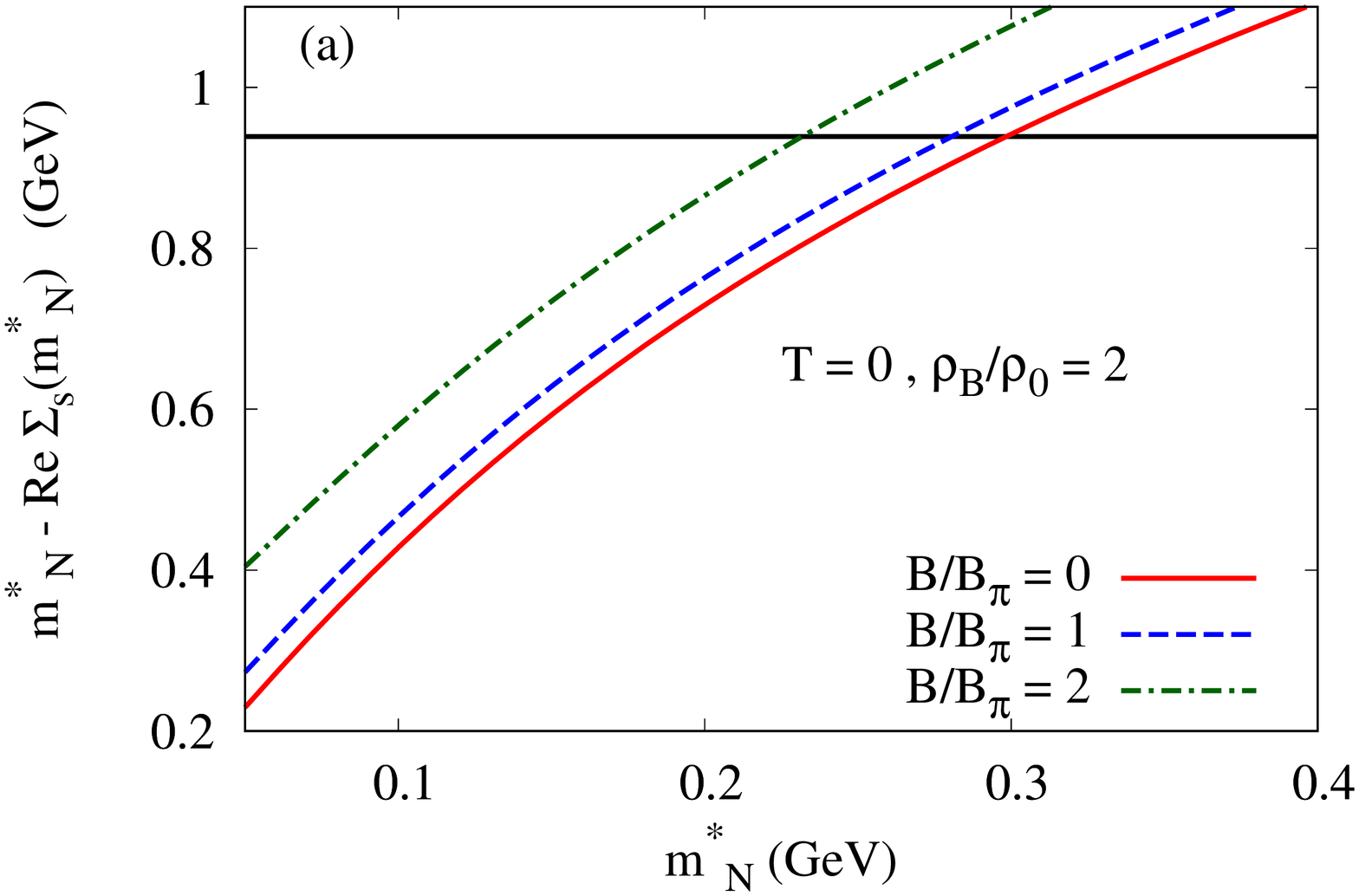}
	\includegraphics[scale=0.3,angle=-90]{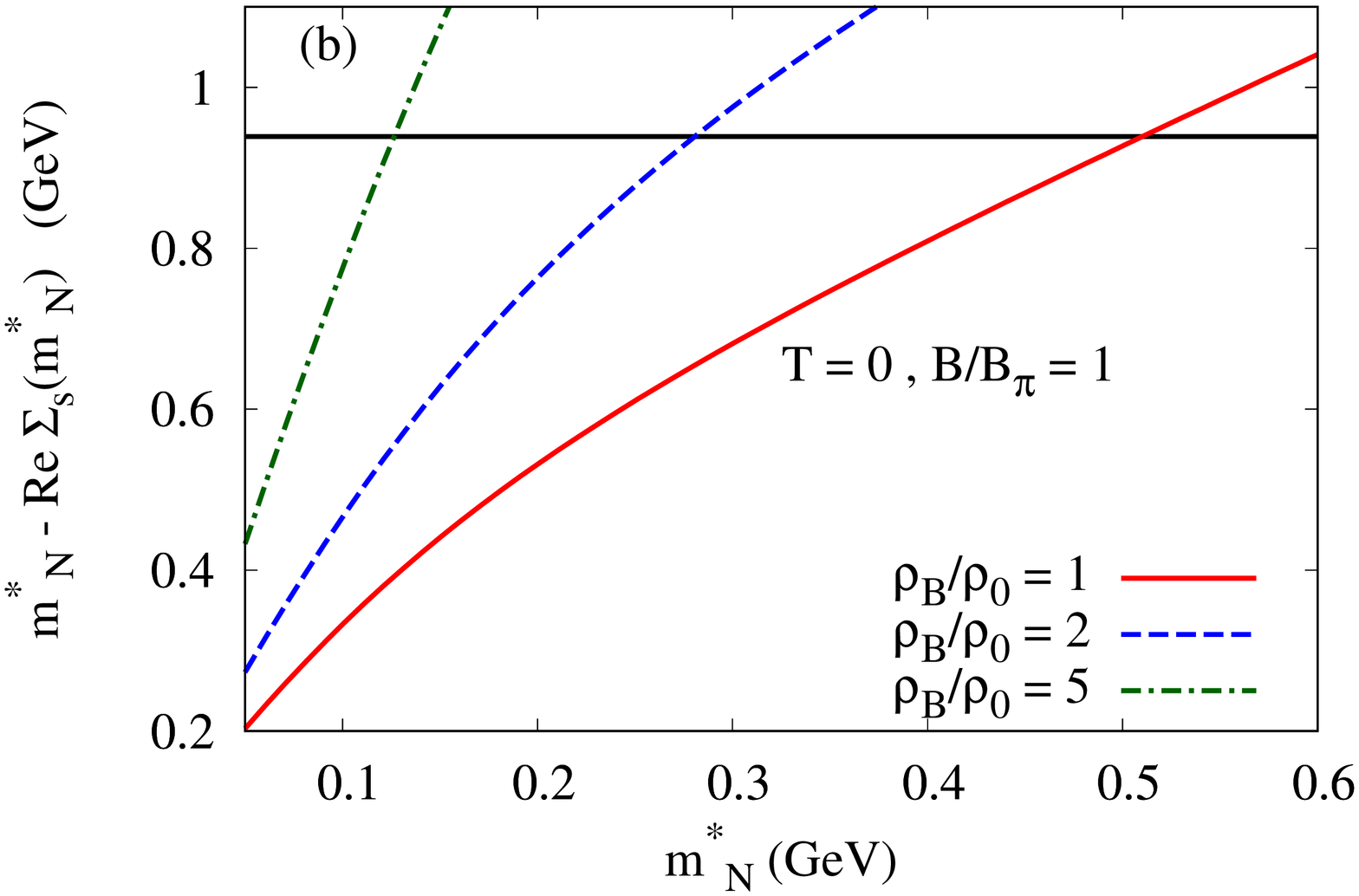}
	\caption{Variation of $m_N^*-\RE\Sigma_s\FB{m_N^*}$ with $m_N^*$ at \textit{zero temperature} for (a) three different values of 
		magnetic field ($B = 0,B_\pi,2B_\pi$) at baryon density $\rho_\text{B}=2\rho_0$ (b) three different values of 
		baryon density ($\rho_B = \rho_0,2\rho_0,5\rho_0$) at magnetic field ($B=B_\pi$). Here $|e|B_\pi = m_\pi^2 = 0.0196~\text{GeV}^2$ 
		and $\rho_0=0.16~\text{fm}^{-3}$. The horizontal black solid line corresponds to $m_N^*=m_N=939$ MeV.}
	\label{fig.sigma.s}
\end{figure}
\begin{figure}[h]
	\includegraphics[scale=0.33,angle=-90]{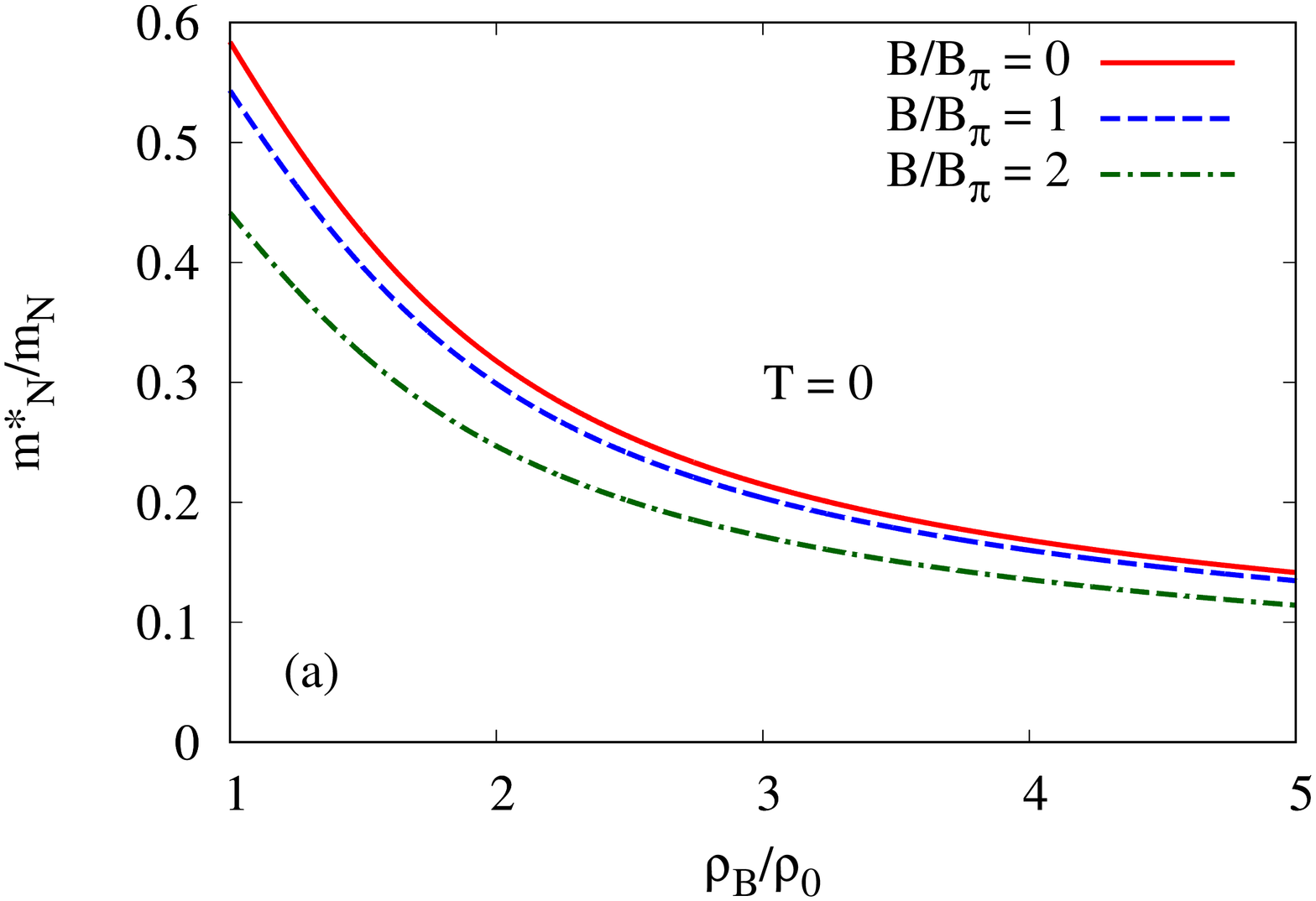} 
	\includegraphics[scale=0.33,angle=-90]{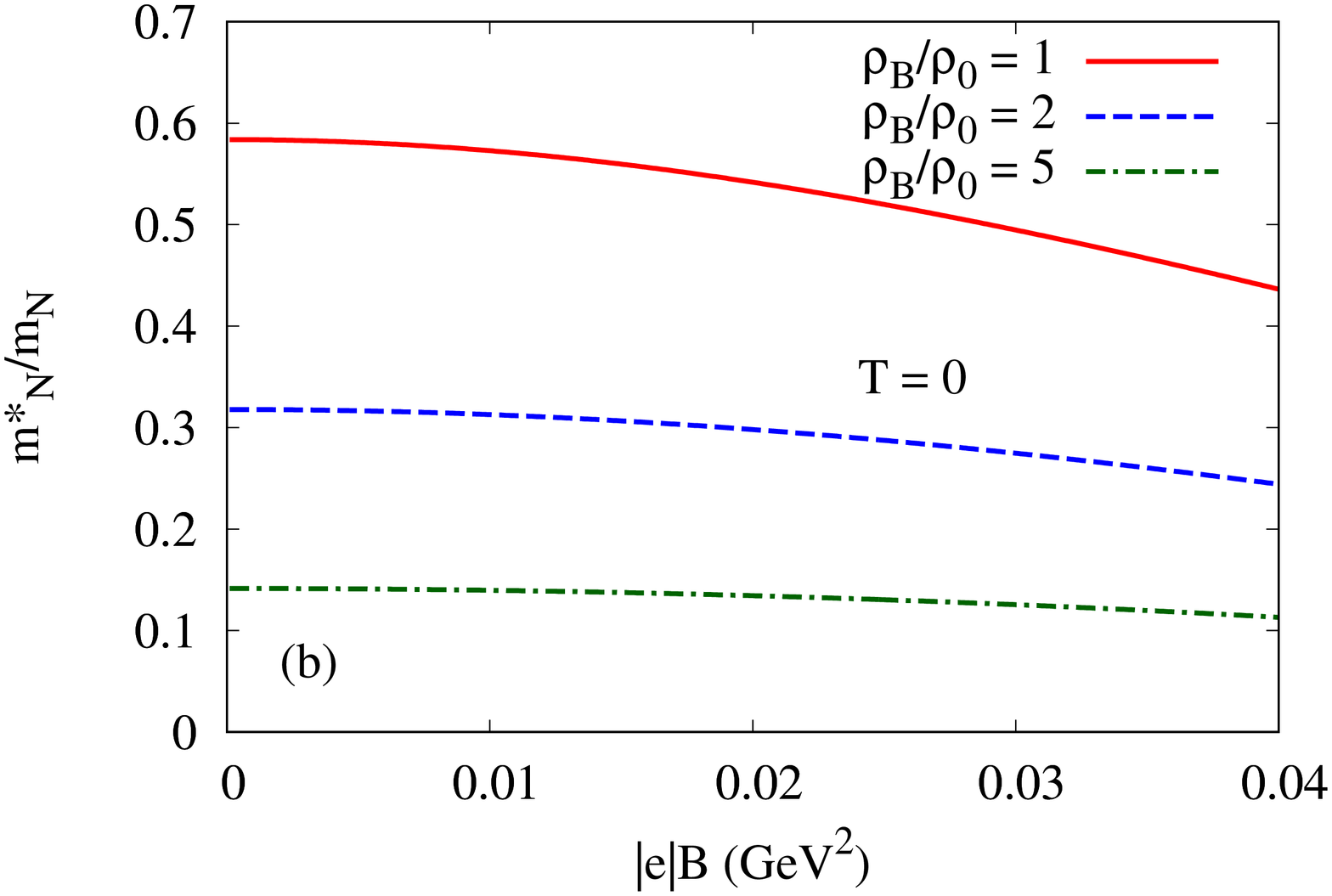} \\
	\includegraphics[scale=0.33,angle=-90]{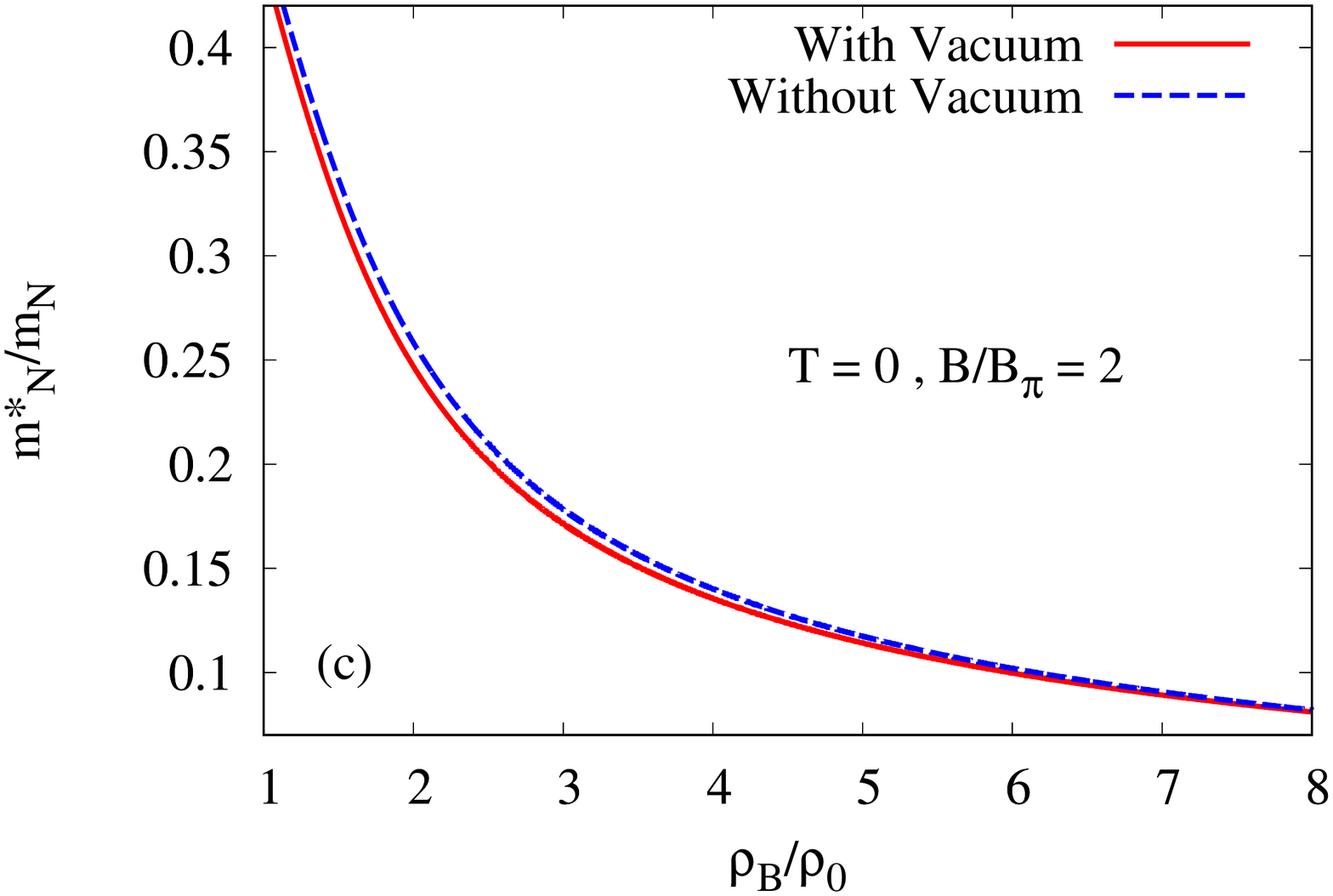} 
	\caption{ Variation of effective mass of nucleon at \textit{zero temperature} 
		(a) with baryon density for	three different values of magnetic field ($B = 0,B_\pi,2B_\pi$). The horizontal axis starts at $\rho_B = 0.1 \rho_0$.
		(b) with magnetic field  for three different values of baryon density ($\rho_B = \rho_0,2\rho_0,5\rho_0$).  
		Here $|e|B_\pi = m_\pi^2 = 0.0196~\text{GeV}^2$ and $\rho_0=0.16~\text{fm}^{-3}$.
	 (c) At $B=2B_\pi$, the variation of the  effective nucleon mass   with baryon density is compared  with the case where the vacuum contribution is ignored.  }
	\label{fig.mstar.t0}
\end{figure}

\begin{figure}[h]
	\includegraphics[scale=0.33,angle=-90]{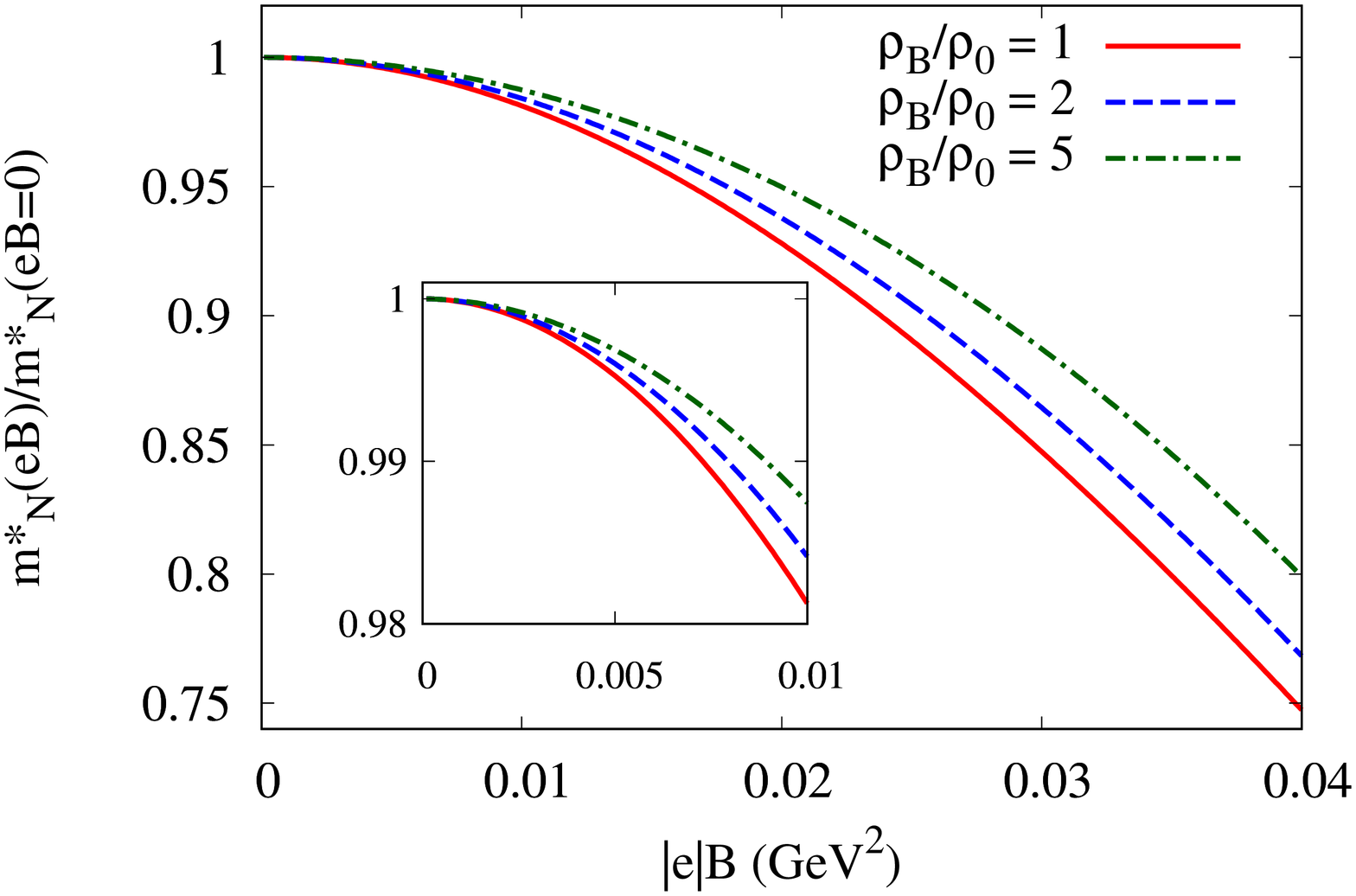} 
	\caption{   At T=0, the ratio of effective mass $m^\ast$ at non-zero $eB$ and  at zero $eB$ is plotted  as a function of $eB$ for three different 
	values of baryon density ($\rho_B = \rho_0,2\rho_0,5\rho_0$). 
	The inset plot shows the low $eB$ region upto $eB=0.01$ GeV$^2$ relevant for neutron star/magnetar case. 
		Here  $\rho_0=0.16~\text{fm}^{-3}$. }
	\label{fig.mstar.t0p}
\end{figure}

Let us now proceed to the  study of  nucleon effective mass  in presence of 
external magnetic field at \textit{at finite baryon density} and \textit{zero temperature}. 
As can be seen from Eqs.~(\ref{eq.sigmavac})-(\ref{eq.sigma.spn}), 
the scalar self energy $\Sigma_s$ is functions of magnetic field $B$ and baryon chemical potential 
$\mu_\text{B}$ of the medium. It is customary to use total baryon density 
$\rho_\text{B}$ instead of $\mu_\text{B}$ where
\begin{eqnarray}
\rho_\text{B} = 4\int\frac{d^3p}{\FB{2\pi}^3}\Theta\FB{\mu_\text{B}-\sqrt{\MB{\vec{p}}^2+m_N^{*2}}}
= \FB{\frac{2}{3\pi^2}}\TB{\mu_\text{B}^2-m_N^{*2}\frac{}{}}^{3/2}~.
\label{eq.rho}
\end{eqnarray}
Inverting the above equation, we get the baryon chemical potential in terms of the baryon density as 
\begin{eqnarray}
\mu_\text{B} = \sqrt{\FB{\frac{3\pi^2}{2}\rho_\text{B}}^{2/3} + m_N^{*2}}~.
\label{eq.mub}
\end{eqnarray}
We have expressed the strength of the magnetic field $B$ with respect to the pion mass scale ($B_\pi$) defined as 
\begin{eqnarray}
\MB{e}B_\pi = m_\pi^2 = 0.0196 \text{ GeV}^2.
\end{eqnarray}
Similarly the total baryon density $\rho_B$ is expressed with respect to the normal nuclear matter 
density $\rho_0 = 0.16$ fm$^{-3}$.

Since we will be solving the transcendental Eq.~(\ref{eq.trans}), we first plot $m_N^*-\RE\Sigma_s\FB{m_N^*}$ as 
a function of $m_N^*$ in Fig.~\ref{fig.sigma.s}.
Fig.~\ref{fig.sigma.s}-(a) depicts the variation of this quantity at three 
different values of magnetic field ($B/B_\pi=$ 0, 1 and 2) with baryon density $\rho_B =2\rho_0$ 
whereas Fig.~\ref{fig.sigma.s}-(b) shows its variation at three different values of total baryon density 
($\rho_B/\rho_0=$ 1, 2 and 3) with magnetic field $B=B_\pi$. The intersections of this graphs with the horizontal line 
corresponding to $m_N^*=m_N=939$ MeV represent the solutions of Eq.~(\ref{eq.trans}). We  notice from these 
figures that $\RE\Sigma\FB{m_N^*}$ is always less than zero and it monotonically decreases as we increase $m_N^*$. 
Also for a particular value of $m_N^*$, $\RE\Sigma\FB{m_N^*}$ decreases with the increase of $B$ and $\rho_B$. 
In Fig.~\ref{fig.mstar.t0}-(a), the  variation of the effective nucleon mass with baryon density has been shown
at three different values of magnetic field ($B = 0,B_\pi,2B_\pi$). As can be seen from the figure, $m_N^*/m_N$ 
decreases  with the increase of $\rho_\text{B}$ and becomes less than  $0.5$ at $\rho_\text{B}=2\rho_0$.
It can be checked that the contribution from the first term within the  square brackets in Eq.~(\ref{eq.sigma.spn}) 
plays the dominant role in determining the $\rho_\text{B}$ as well as the  $eB$ dependences of the effective mass whereas 
 the net contribution from all the other terms in $\Sigma_s^\text{(medium)}$ and   $\Sigma_s^\text{(vacuum)}$ (see  Eq.~(\ref{eq.sigmavac})) 
 remains sub-leading throughout. Also, it is clear from Fig.~\ref{fig.mstar.t0}-(a) that,  with the increase of $|e|B$, the effective mass
 decreases and the effect of the external magnetic field 
is more at a lower $\rho_\text{B}$ region.   At very high $\rho_\text{B} (\gtrsim 5\rho_0)$ it is expected that 
the effect of $\MB{e}B$ on nucleon effective mass becomes negligible. However, the conclusions based on the 
weak field approximation will not be reliable for arbitrary large or small  densities  as will be discussed later.

In Fig.~\ref{fig.mstar.t0}-(b), the variation of $m_N^*/m_N$ with $\MB{e}B$ is shown at three 
different values of baryon density ($\rho_B = \rho_0,2\rho_0,5\rho_0$). We find a small decrease in effective nucleon 
mass with  $|e|B$.
 In order to observe the effect of the vacuum self energy correction to the effective mass of nucleon, 
	we have compared the density variations of $m_N^*$ with and   without the  vacuum contribution  as shown in Fig.~\ref{fig.mstar.t0}(c). Here 
	the external magnetic field is kept fixed at $B=2B_\pi$. 
	It has been noticed that the effect of vacuum correction is subleading with respect to the medium contribution at 
	non-zero baryon density and the correction to $m_N^*$ due to vacuum self energy remains less than 6\%.
It is also interesting to observe  the relative importance of the external magnetic field on the effective nucleon mass
as shown in Fig.~\ref{fig.mstar.t0p} where 
 the ratio $m_N^*(eB)/m_N^*(eB=0)$  is plotted as a function of $eB$  at three 
different baryon densities( $\rho_B = \rho_0,2\rho_0,5\rho_0$).
It can be noticed that $m_N^*$ decreases by about 25\% at a magnetic field $eB\sim 0.04$ GeV$^2$. The inset plot shows 
the lower $eB$ region upto $eB=0.01$ GeV$^2$ which corresponds to the typical values of  magnetic field expected  inside a neutron star/magnetar. 
At the  maximum  value  $eB=0.01$ GeV$^2$, the effective mass of nucleon is found to be lowered by less than 2\%.

\begin{figure}[h]
	\includegraphics[scale=0.33,angle=-90]{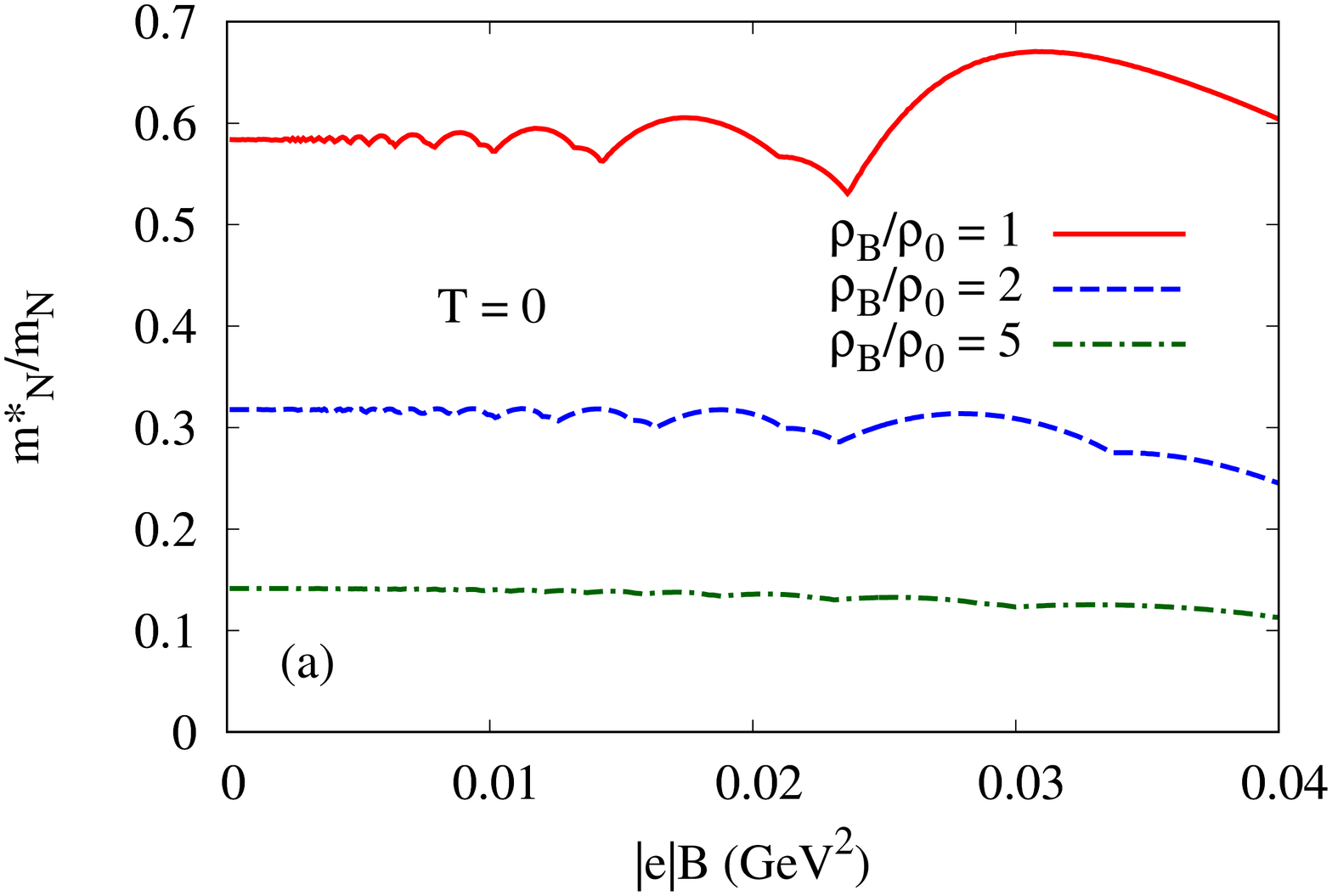} 
	\includegraphics[scale=0.33,angle=-90]{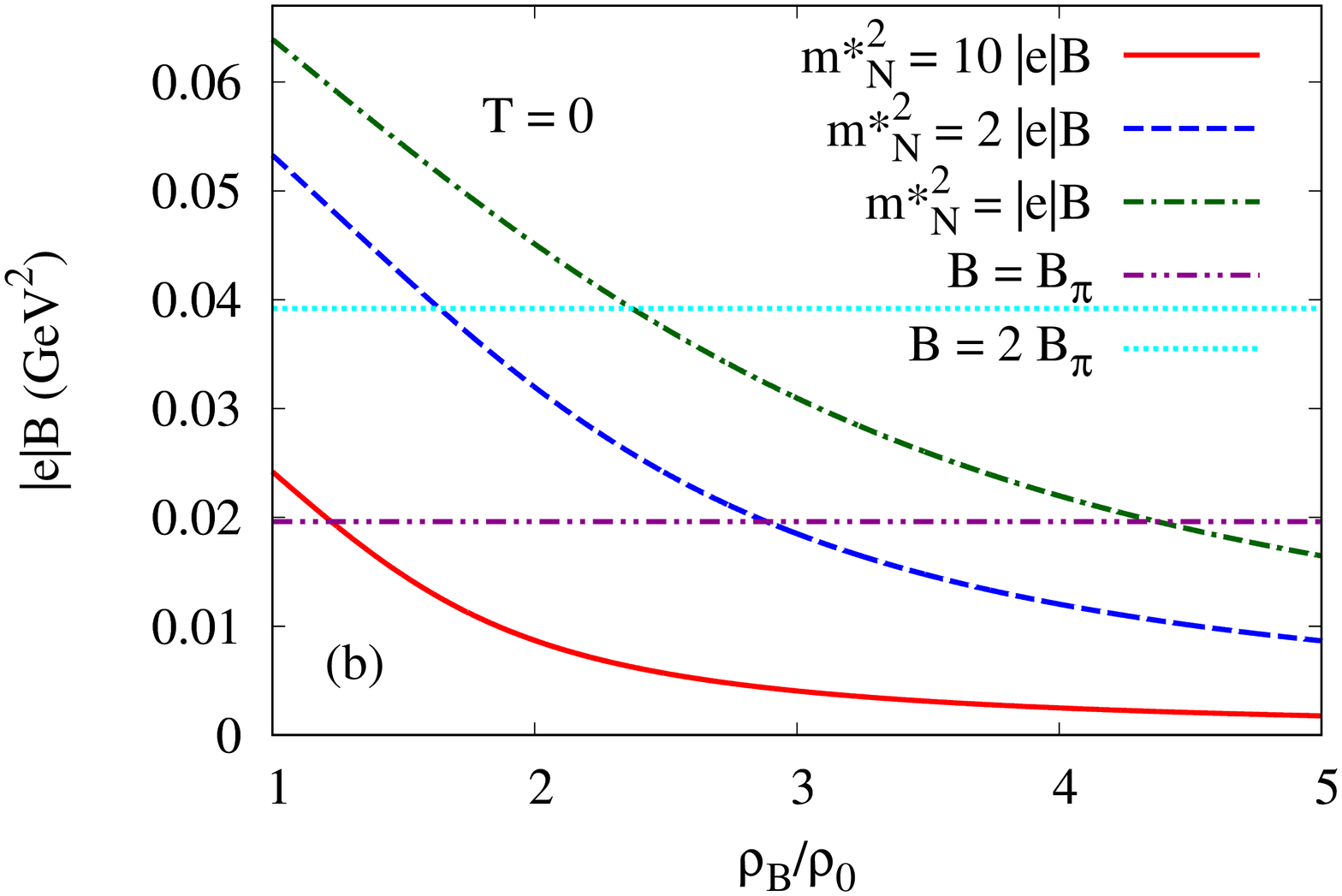} 
	\caption{ Variation of effective mass of nucleon at \textit{zero temperature} 
		(a) with baryon density for	three different values of magnetic field ($B = 0,B_\pi,2B_\pi$). The horizontal axis starts at
		$\rho_B = 0.1 \rho_0$.
		(b) with magnetic field  for three different values of baryon density ($\rho_B = \rho_0,2\rho_0,5\rho_0$).  
		Here $|e|B_\pi = m_\pi^2 = 0.0196~\text{GeV}^2$ and $\rho_0=0.16~\text{fm}^{-3}$.}
	\label{fig.mstar.t00}
\end{figure}

Until now we have considered that under weak field approximation, the modifications  from the  non-vanishing  anomalous magnetic moment 
 arise only through the  effective mass. Moreover, it is assumed that 
the modification in the expression of proton  density  as a summation over 
Landau levels can also  be ignored for  weak external fields. The motivation behind this approximation lies in the fact that with smaller values of external field, the Landau levels
become more and more closely spaced giving rise to a continuum at $eB\rightarrow0$. In that case, the summations that appeared due to the Landau quantization, can be
replaced by the  corresponding momentum  integrals giving rise to exactly similar  expression for proton and neutron density in isospin symmetric matter. 
As a result, the expression of baryon density as  given 
in Eq.~(\ref{eq.rho}) remains to be valid even in presence of $eB$  as long as the external fields are sufficiently  weak to make the 
summation to integral conversion plausible. It is advantageous to use this approximate expression to obtain the effective mass of the nucleons 
as, in this case, $\mu_\text{B}$ can be analytically expressed 
in terms of $\rho_\text{B}$    providing  useful  simplifications in  the numerics.  However, to check the validity of the approximations,  
it is reasonable to  incorporate this magnetic modifications in
the expression for the  net baryon density which now becomes \cite{Aguirre1,Aguirre2} 
	\begin{eqnarray}
	\rho_\text{B} &=& \sum_{s\in\{\pm1\}}^{}\int\frac{d^3p}{\FB{2\pi}^3}\Theta\SB{\mu_\text{B}-\sqrt{p_z^2+\FB{\sqrt{m_N^{*2}-p_\perp^2}-s\kappa_{\text{n}}B}^2}} \nn \\
	&& \hspace{1cm}+ \frac{eB}{(2\pi)^2}\sum_{s\in\{\pm1\}}^{}\sum_{n=0}^{\infty} (1-\delta_0^n\delta_{-1}^s)\int_{-\infty}^{\infty}dp_z
	\Theta\SB{\mu_\text{B}-\sqrt{p_z^2+\FB{\sqrt{m_N^{*2}+2n|e|B}-s\kappa_{\text{p}}B}^2}}.
	\label{eq.rho.eb}
	\end{eqnarray}
	 Performing 
	the momentum integral in the above equation, we obtain
	\begin{eqnarray}
	\rho_\text{B} &=& \sum_{s\in\{\pm1\}}^{} \frac{1}{12\pi^2} \left[3 \pi \mu_\text{B}^2 s\kappa_{\text{n}}B+2 \sqrt{\mu_\text{B} ^2-(m_N^*-s\kappa_{\text{n}}B)^2} \left\{\frac{}{}2 \mu_\text{B} ^2-2 m_N^{*2}+m_N^* s\kappa_{\text{n}}B+(s\kappa_{\text{n}}B)^2\right\} \right. \nn \\
	&&\left.+6 \mu_\text{B} ^2 s\kappa_{\text{n}}B \tan
	^{-1}\left\{\frac{s\kappa_{\text{n}}B-m_N^*}{\sqrt{\mu_\text{B} ^2-(m_N^*-s\kappa_{\text{n}}B)^2}}\right\}\right] \nn \\
	&& + \frac{eB}{2\pi^2}\sum_{s\in\{\pm1\}}^{}\sum_{n=0}^{n_\text{max}}(1-\delta_0^n\delta_{-1}^s)\sqrt{\mu_\text{B}^2-\FB{\sqrt{m_N^{*2}+2n|e|B}-s\kappa_{\text{p}}B}^2}
	\end{eqnarray}
	where, $n_\text{max}=\TB{\frac{(\mu_\text{B}+s\kappa_{\text{p}}B)^2-m_N^{*2}}{2|e|B}}$ in which $\TB{x}=$ greatest integer less 
	than or equal to $x$.
	The above equation can not be inverted analytically in order to express $\mu_B$ as a function of $\rho_B$ which 
	was possible for $eB=0$ case (see  Eq.~(\ref{eq.mub})). Thus we invert the  equation  numerically 
	to obtain $\mu_B=\mu_B(\rho_B,eB)$. Using the above modified $\rho_B$, we have re-plotted the effective mass variation with the external field for 
	the same set of densities $\rho_B = \rho_0,2\rho_0$ and $5\rho_0$ as shown in Fig.~\ref{fig.mstar.t00}. The  oscillating  
	behaviour  is consistent with  Ref.~\cite{Haber:2014ula}. Comparison with Fig.\ref{fig.mstar.t0}(b) suggests that the usual baryon
	 density expression provides the average qualitative behaviour reasonably well even in presence of external magnetic field as long as the background field 
	 strength is small and the agreement is more pronounced in higher density regime.  However, going to arbitrary large densities is restricted by the assumption 
	 of weak field expansion of the 
	 propagator  which demands  the external  $eB$ to be much smaller than  $m_N^{\ast 2}$. Now, apart from the external magnetic field, this effective mass depends on density as well and more importantly, 
	 the dependence is of decreasing nature. Thus, even if one starts with a constant $eB$ much lower than  $m_N^{\ast 2}$, the decreasing 
	 trend of $m_N^{\ast}$ with density invalidates this basic weak field assumption at some higher $\rho_\text{B}$ value for which $m_N^{\ast 2}$ becomes comparable with 
	 the constant $eB$ used. To estimate this density value, we fix the maximum possible value of $eB$ to be considered as a fraction times  $m_N^{\ast 2}$ where 
	 the fraction is chosen to be 0.5 and 0.1. The corresponding variation with respect to $\rho_\text{B}$ are shown in Fig.\ref{fig.mstar.t00}(b) where the case 
	 $eB=m_N^{\ast 2}$ is also plotted for comparison. Each of these curves in fact serves the purpose of a boundary and for a given value of 
	 $\rho_\text{B}$, only those  $eB$  values are allowed which lie below it.  The horizontal lines denote the constant magnetic field values used in this work. 
	 It is clear from the figure that, once we have chosen the maximum $eB$ curve( say  $eB=0.5m_N^{\ast 2}$ curve), its intersection with each horizontal lines provides
	 the maximum density ( i.e around $3\rho_0$ for $B=B_\pi$ and around $1.8\rho_0$ for $B=2B_\pi$) up to which the $eB$ value corresponding to that line 
	 can be considered as `weak'. 

\begin{figure}[h]
	\includegraphics[scale=0.3,angle=-90]{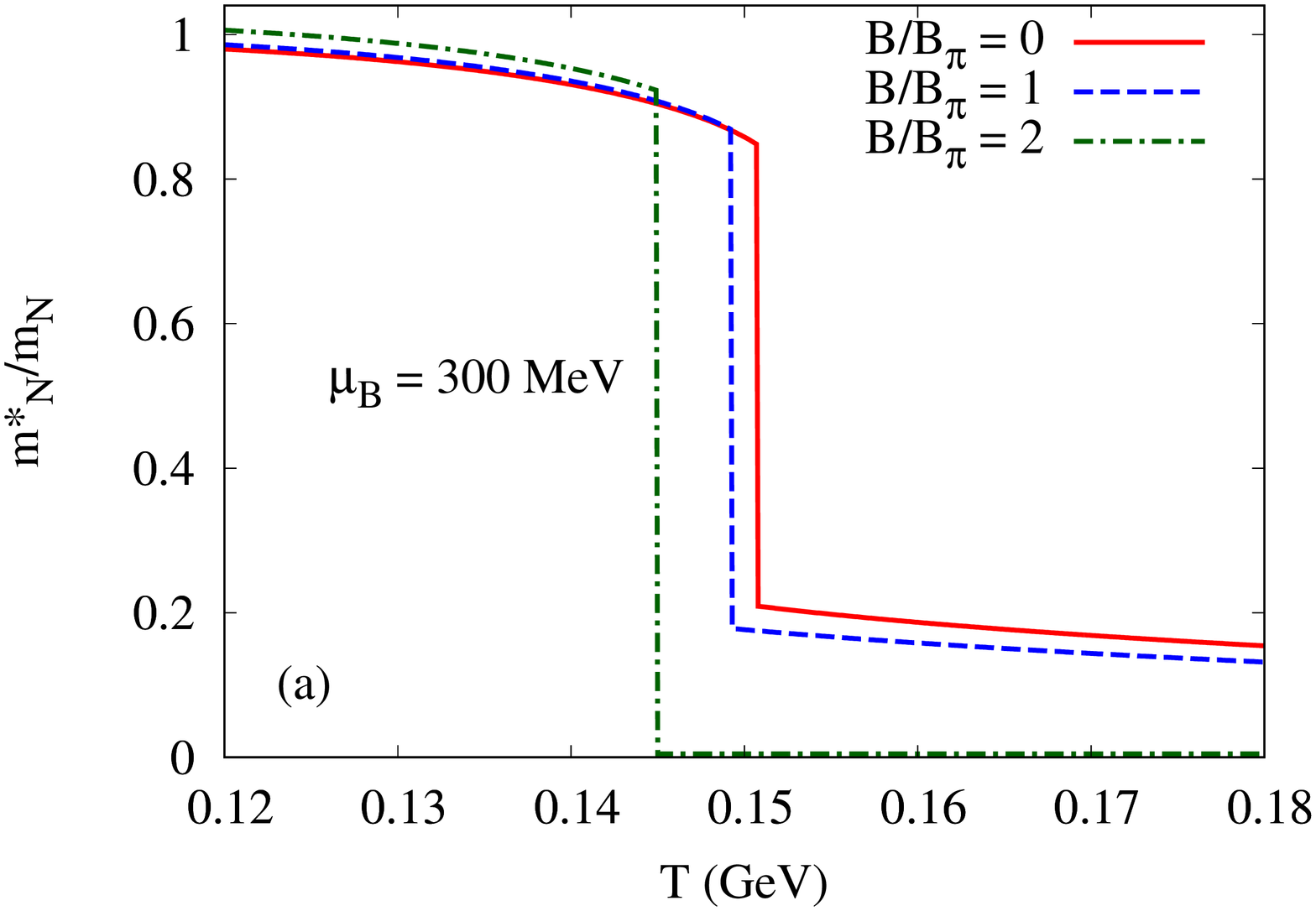}
	\includegraphics[scale=0.3,angle=-90]{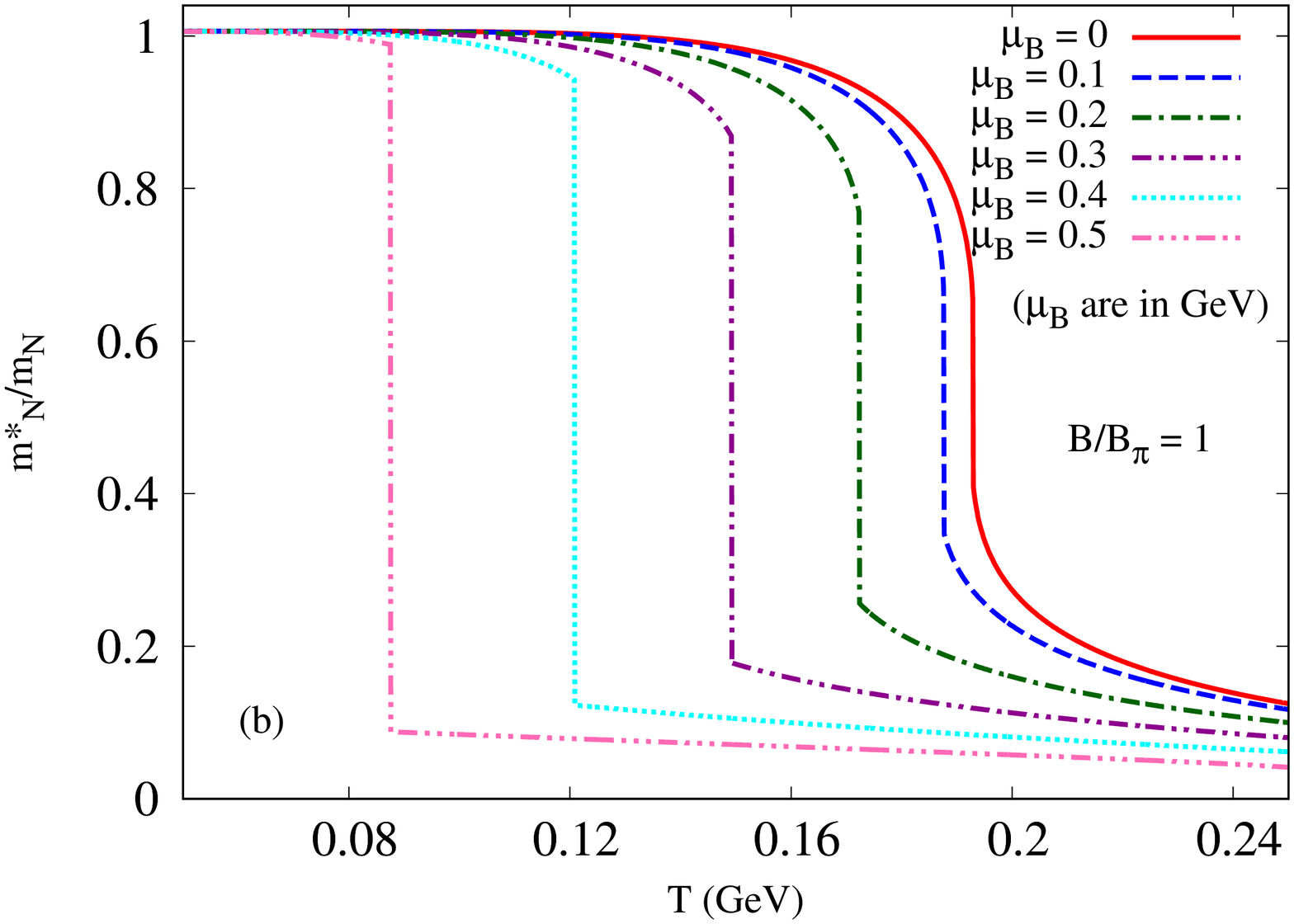}
	\includegraphics[scale=0.3,angle=-90]{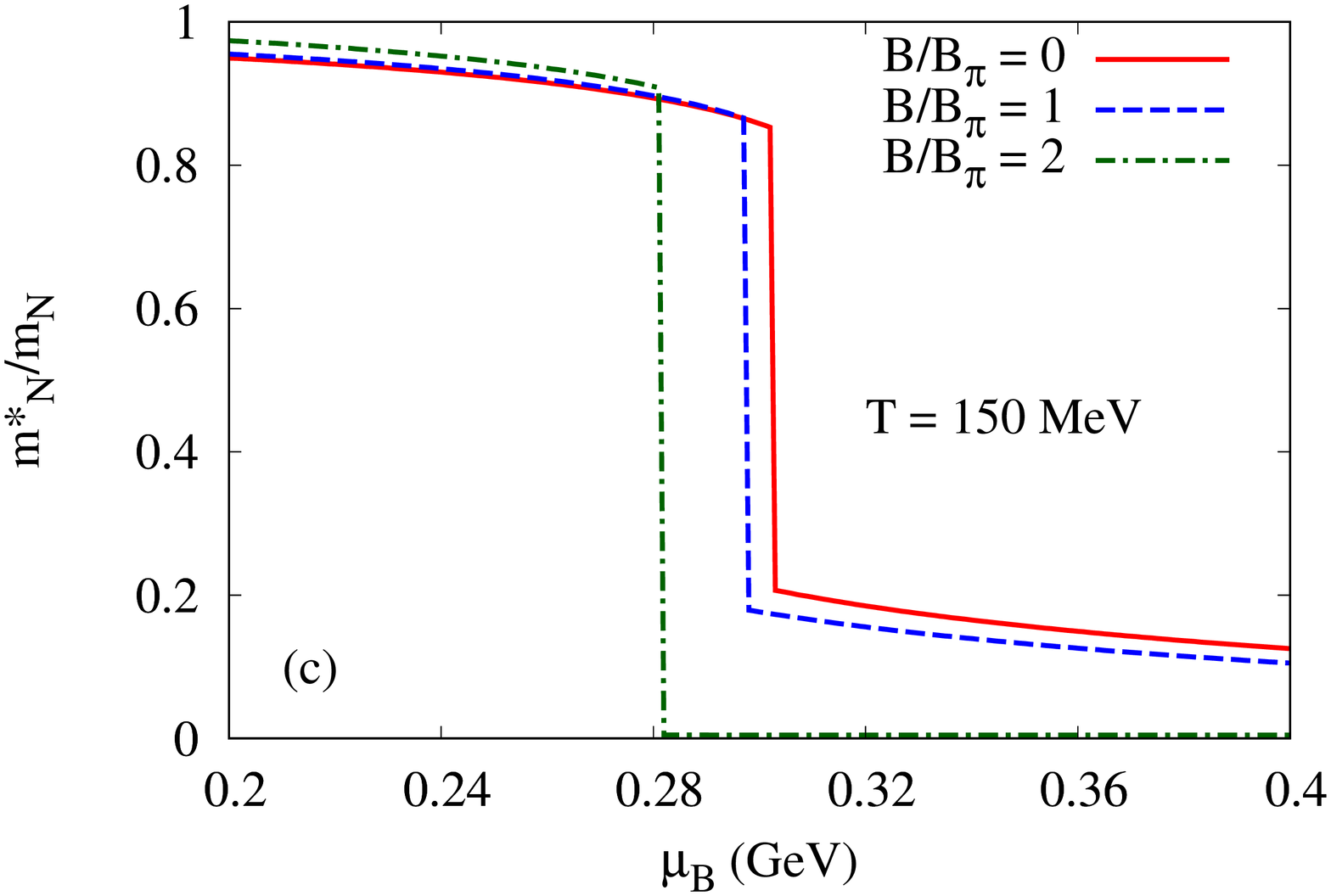}
	\includegraphics[scale=0.3,angle=-90]{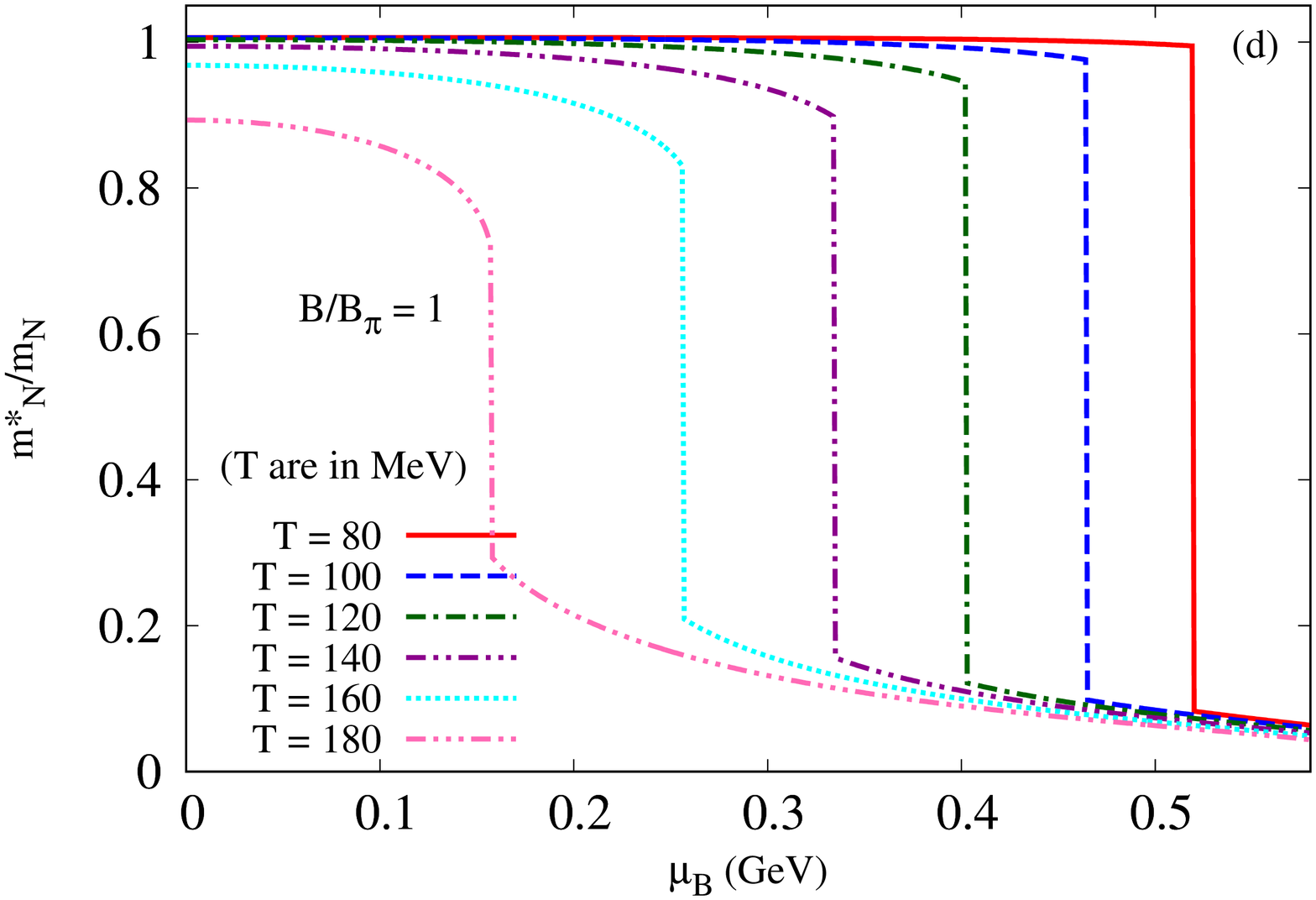}
	\caption{Variation of effective mass of nucleon with temperature (a) at $\mu_\text{B}$=300 MeV 
		for three different values $B$ ($0,B_\pi$ and $2B_\pi$) (b) at $B=B_\pi$ for six different values 
		$\mu_\text{B}$ (0, 100, 200, 300, 400 and 500 MeV). 
		Variation of effective mass of nucleon with baryon chemical potential (a) at $T$=150 MeV 
		for three different values of magnetic field ($B = 0,B_\pi,2B_\pi$) (b) at $B=B_\pi$ for six different value of 
		$T$ = 80, 100, 120, 140, 160 and 180 MeV.
		Here $|e|B_\pi = m_\pi^2 = 0.0196~\text{GeV}^2$.  }
	\label{fig.mstar.t}
\end{figure}

\begin{figure}[h]
	\includegraphics[scale=0.3,angle=-90]{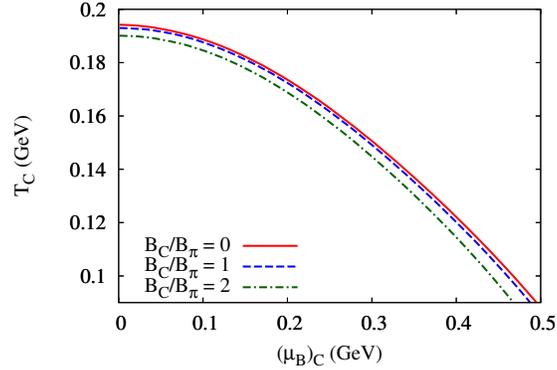}
	\caption{Phase diagram for vacuum to nuclear medium phase transition in Walecka model for three 
		different values of $B$ ($0,B_\pi$ and $2B_\pi$). }
	\label{fig.phase}
\end{figure}
\begin{figure}[h]
	\includegraphics[scale=0.3,angle=-90]{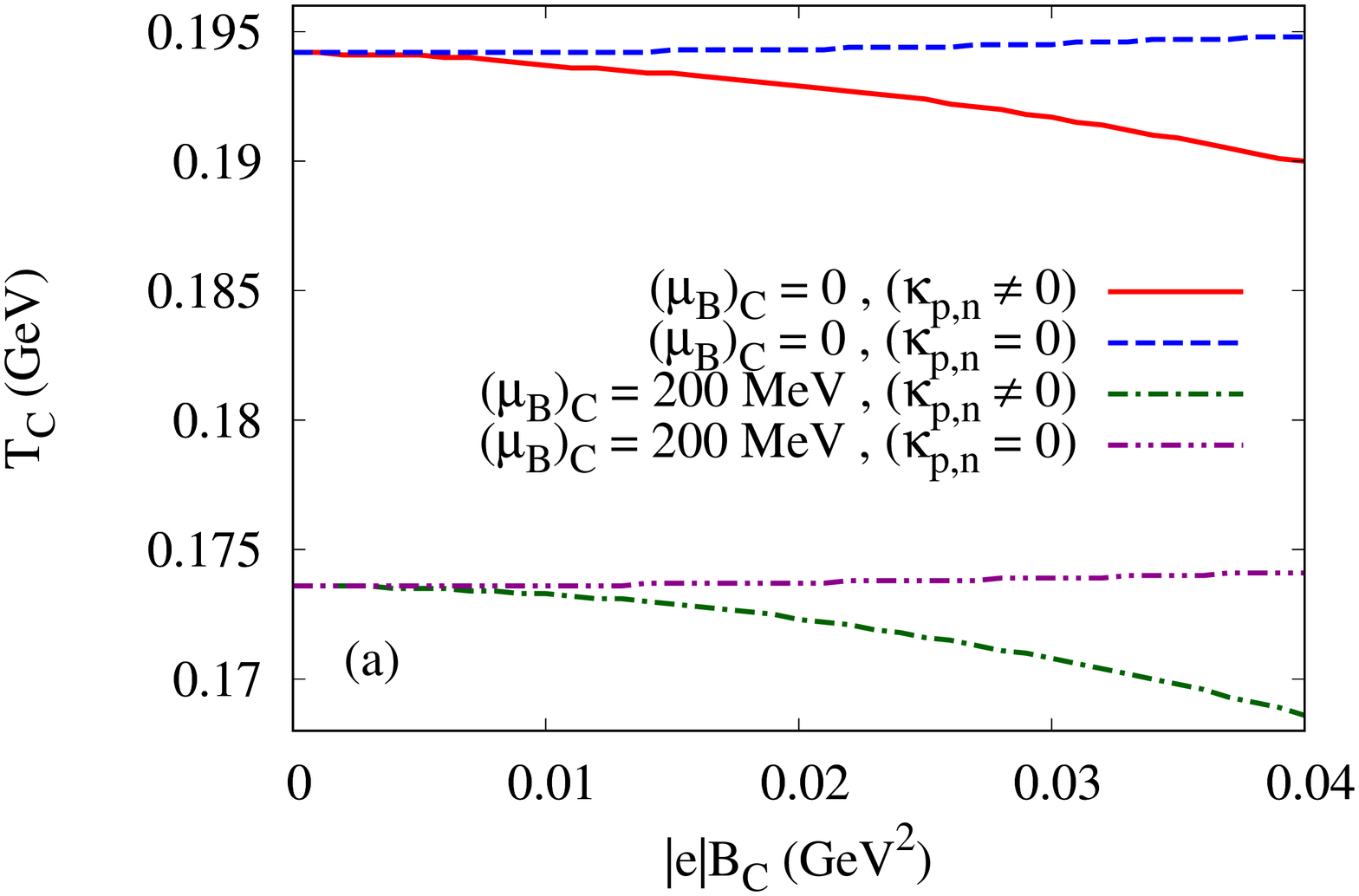}
	\includegraphics[scale=0.3,angle=-90]{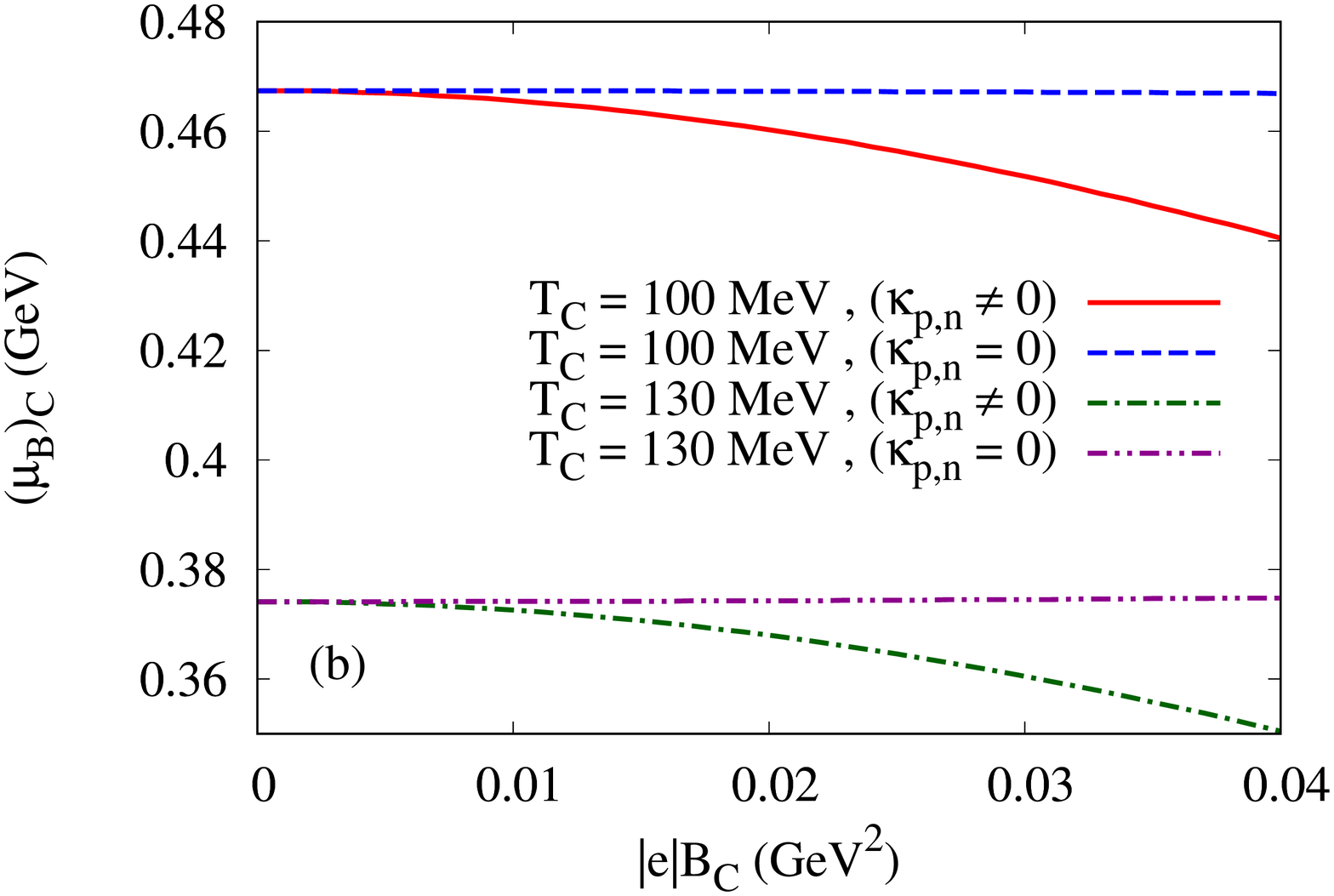}
	\caption{(a) Variation of transition temperature with magnetic field at two different values of $\mu_\text{B}$ (0 and 200 MeV).
		(b) Variation of transition baryon chemical potential with magnetic field at two different values 
		of $T$ (100 and 130 MeV). Cases with and without the ANM of nucleons are shown separately.}
	\label{fig.tcmubc}
\end{figure}

We now turn on the \textit{temperature} and study the variation of $m_N^*/m_N$ with temperature and baryon chemical potential 
in Fig.~\ref{fig.mstar.t}. Fig.~\ref{fig.mstar.t}-(a) depicts the variation of $m_N^*/m_N$ with $T$ at at $\mu_\text{B}$=300 MeV 
and at three different values $B$ ($0,B_\pi$ and $2B_\pi$) whereas Fig.~\ref{fig.mstar.t}-(b) shows its variation 
at $B=B_\pi$ and at six different values $\mu_\text{B}$ (0, 100, 200, 300, 400 and 500 MeV). As can be seen from the figure, 
that the effective nucleon mass suffers a sudden decrease at a particular temperature corresponding to the 
vacuum to nuclear medium phase transition~\cite{Negele:1986bp,Haber:2014ula}. 
We call this transition temperature as $T_C$ which we calculate numerically from the slope of of these plots. 
As can be seen from Fig.~\ref{fig.mstar.t}-(a), $T_C$ decreases with the increase of $B$, which may be identified as 
IMC in Walecka model.
In Fig.~\ref{fig.mstar.t}-(b), we observe that $T_C$ decreases with the increase of $\mu_\text{B}$.
The corresponding variation of $m_N^*/m_N$ with $\mu_\text{B}$ is shown in Fig.~\ref{fig.mstar.t}-(b) and (c). 
Analogous to the upper panels, we see the phase transition at a particular $\mu_\text{B}$ and we call this transition 
chemical potential as $\FB{\mu_\text{B}}_C$. As can be seen in the graphs, $\FB{\mu_\text{B}}_C$ decreases with 
the increase in $B$ and $T$.

The behaviour of $T_C$ and $\FB{\mu_\text{B}}_C$ at different $B$ can be seen in Fig.~\ref{fig.phase}, where, 
we have presented the phase diagram for the vacuum to nuclear medium phase transition at 
three different values of $B$ ($0,B_\pi$ and $2B_\pi$). With the increase in $\FB{\mu_\text{B}}_C$, $T_C$ decreases and vice-versa. 
Also, with the increase in $B$, the phase boundary in this $T-\mu_\text{B}$ plane moves towards lower 
values of $T$ and $\mu_\text{B}$ showing IMC.

We conclude this section by presenting the variation of $T_C$ and $\FB{\mu_\text{B}}_C$ with external magnetic field in 
Fig.~\ref{fig.tcmubc}. Fig.~\ref{fig.tcmubc}-(a) shows the variation of $T_C$ with $|e|B$ at two different 
values of $\mu_\text{B}$ (0 and 200 MeV) whereas Fig.~\ref{fig.tcmubc}-(b) shows the corresponding variation at  
at two different values of $T$ (100 and 130 MeV). As already discussed, both the $T_C$ and $\FB{\mu_\text{B}}_C$ 
decreases with the increase in $B$ characterizing the IMC effect.
However, once the anomalous magnetic moment is ignored, $T_C$ as well as $\FB{\mu_\text{B}}_C$ can be observed to slowly increase with the external magnetic 
field showing MC as expected \cite{Haber:2014ula}.


\section{Summary}
\label{summary}
In this article we have used  
the  Walecka model to study the vacuum to nuclear matter phase transition in presence of a weak and constant background magnetic field within mean field approximation. 
In case of weak magnetic 
field, the nucleon propagators are derived as  a series in powers of $qB$ and $\kappa B$ where $q$ and $\kappa$ represents the charge and the anomalous magnetic moment 
of the nucleons.  The effective mass of the nucleon ($m_N^*$) is obtained from  the pole of the  nucleon propagator self-consistently. At zero temperature and zero density, 
the incorporation of anomalous magnetic moment is shown to favour the effective  mass enhancement with the external magnetic field. The functional dependence of  $m_N^*$ on
the background field is extended to the case of  non-zero nuclear density and further extended to the finite temperature regime. It is observed that  in the case of vanishing 
temperature within dense nuclear medium, the  effective  mass  decreases with the background  magnetic field and this trend is shown to  survive in case of
non-zero temperature as well. Moreover, there exists  a particular temperature (denoted by $T_C$ in the text) for which the effective nucleon mass suffers a 
sudden decrease  corresponding to  the vacuum to nuclear medium phase transition.  It has been  shown that this critical temperature  decreases 
with the increase of $B$ which can  be identified as inverse magnetic catalysis  in Walecka model whereas  the opposite behaviour is obtained in case of 
vanishing  magnetic moment. Thus, it can be inferred that  in presence of external magnetic field, the  anomalous magnetic moment of the nucleons  plays a crucial role 
in characterizing the nature of vacuum to nuclear matter transition at finite temperature and density.It should be mentioned here that  
 Haber et.al \cite{Haber:2014ula} had speculated that the incorporation of anomalous magnetic moment could 
counteract the effect of magnetic catalysis \cite{haber37}. Our study not only supports the speculation but also concludes that 
the effect is significant enough to alter the qualitative behaviour of the nucleon effective mass even in weak magnetic field regime.  However, it should 
be noted here that the weak field approximation actually restricts the regime  of validity of the present study as discussed in detail in the text. The maximum 
value of the external magnetic field used in the present study is taken to be 0.04 GeV$^2$ and it has been argued to be considered as `weak' only up to density 
1.8 $\rho_0$ where the assumption of `weakness' is fixed by the condition that the chosen external field has to  remain less than 
50\% of the effective mass.  One should also notice that in case of Walecka model, MC or IMC can only be seen indirectly.   Similar studies in 
extended linear sigma model might be interesting as in that case the possibility of (approximate) chiral symmetry restoration is incorporated within the 
model framework. However, we should also mention that in case of zero magnetic moment, only the quantitative difference in the behaviour of the 
effective mass is found to be attributed to the presence of the chiral partners \cite{Haber:2014ula} whereas the qualitative behaviour
which has been the main interest throughout  the article seems to show model independence. Before applying  the present result to obtain the characteristics
of compact stars such as mass radius relationship or the equation of state, beta equilibrium and charge neutrality conditions have to be properly 
incorporated which lies beyond the scope of the present study.

\section*{Acknowledgement}
Snigdha Ghosh acknowledges Center for Nuclear Theory, Variable Energy Cyclotron Centre and Indian Institute of 
Technology Gandhinagar for support.
%


\appendix

\section{Calculation of $\Sigma_s^{\text{(vacuum)}}$}\label{appendix.sigma.s.vacuum}

We have from Eq.~(\ref{eq.sigma.s.B}),
\begin{eqnarray}
\Sigma_s^{\text{(vacuum)}} = \FB{\frac{g^2_{\sigma NN}}{m_\sigma^2}}\RE ~i\int\frac{d^dp}{\FB{2\pi}^d}
\hat{T}\FB{p,m_N^*,m_1}\left.\frac{1}{p^2-m_1^2+i\epsilon}\right|_{m_1=m_N^*,d\rightarrow4} \label{eq.sigma.s.B.2}
\end{eqnarray}
In order to perform the $d^4p$ integration, we use the following identities~\cite{Peskin:1995ev}
\begin{eqnarray}
\int\frac{d^dp}{\FB{2\pi}^d}\FB{\frac{1}{p^2-\Delta}} &=& \frac{-i}{\FB{4\pi}^{d/2}}\Gamma\FB{1-\frac{d}{2}}\FB{\frac{1}{\Delta}}^{1-d/2}  \\
\int\frac{d^dp}{\FB{2\pi}^d}\FB{\frac{p_\perp^2}{p^2-\Delta}} &=& \frac{i}{\FB{4\pi}^{d/2}}\FB{\frac{d}{4}}\Gamma\FB{-\frac{d}{2}}\FB{\frac{1}{\Delta}}^{-d/2} \\
\int\frac{d^dp}{\FB{2\pi}^d}\FB{\frac{p_\parallel^2}{p^2-\Delta}} &=& \frac{i}{\FB{4\pi}^{d/2}}\FB{\frac{d}{4}}\Gamma\FB{-\frac{d}{2}}\FB{\frac{1}{\Delta}}^{-d/2} \\
\int\frac{d^dp}{\FB{2\pi}^d}\FB{\frac{p^2}{p^2-\Delta}} &=& \frac{i}{\FB{4\pi}^{d/2}}\FB{\frac{d}{2}}\Gamma\FB{-\frac{d}{2}}\FB{\frac{1}{\Delta}}^{-d/2}
\end{eqnarray}
so that, Eq.~(\ref{eq.sigma.s.B.2}) will become
\begin{eqnarray}
\Sigma_s^{\text{(vacuum)}} &=& \RE\Sigma_s^\text{(pure vacuum)} + \Sigma_s^\text{(divergent)} + \Sigma_s^\text{(regular)}
\end{eqnarray}
where $\RE\Sigma_s^\text{(pure vacuum)}$ is the ultra-violate divergent pure vacuum contribution given in Eq.~\ref{eq.sigma.s.purevacuum} and
\begin{eqnarray}
\Sigma_s^{\text{(divergent)}} &=& -\FB{\frac{g^2_{\sigma NN}}{4\pi^2m_\sigma^2}}
\SB{\FB{\kappa_{\text{p}}B}^2m_N^*+\FB{\kappa_{\text{n}}B}^2m_N^* \frac{}{}
+\frac{}{}\FB{\MB{e}B}\FB{\kappa_{\text{p}}B}}\left.\Gamma\FB{2-\frac{d}{2}}\FB{\frac{1}{m_N^{*2}}}^{2-d/2}
\right|_{d\rightarrow4} \\
\Sigma_s^{\text{(regular)}} &=& \FB{\frac{g^2_{\sigma NN}}{4\pi^2m_\sigma^2}}
\TB{\frac{\FB{eB}^2}{3m_N^*}+\frac{1}{2}\SB{\FB{\kappa_{\text{p}}B}^2m_N^*+\FB{\kappa_{\text{n}}B}^2m_N^* \frac{}{}
+\frac{}{}\FB{\MB{e}B}\FB{\kappa_{\text{p}}B}} }~.
\end{eqnarray}
In this case also, we will neglect the pure vacuum contribution $\RE\Sigma_s^{(\text{pure vacuum})}$ which is equivalent to use 
the MFT. We now extract the divergence of $\Sigma_s^{\text{(divergent)}}$ from the pole of the Gamma function 
and use $\overline{\text{MS}}$ scheme to obtain, 
\begin{eqnarray}
\Sigma_s^{\text{(divergent)}} &=& \FB{\frac{g^2_{\sigma NN}}{4\pi^2m_\sigma^2}}
\SB{\FB{\kappa_{\text{p}}B}^2m_N^*+\FB{\kappa_{\text{n}}B}^2m_N^* \frac{}{}
	+\frac{}{}\FB{\MB{e}B}\FB{\kappa_{\text{p}}B}}\ln\FB{\frac{m_N^{*2}}{\Lambda}}
\end{eqnarray}
where $\Lambda$ is a scale of dimension GeV$^2$. Its value is fixed from the 
condition $\Sigma_s^\text{(divergent)}\FB{m_N^*=m_N} = 0$, which gives $\Lambda=m_N^2$. So the final expression of 
$\Sigma_s^{\text{(vacuum)}}$ becomes
\begin{eqnarray}
\Sigma_s^{\text{(vacuum)}} &=& \FB{\frac{g^2_{\sigma NN}}{4\pi^2m_\sigma^2}}
\TB{\frac{\FB{eB}^2}{3m_N^*}+\SB{\FB{\kappa_{\text{p}}B}^2m_N^*+\FB{\kappa_{\text{n}}B}^2m_N^* 
+\FB{\MB{e}B}\FB{\kappa_{\text{p}}B}}\SB{\frac{1}{2}+2\ln\FB{\frac{m_N^*}{m_N} } }}~ \label{eq.sigmavac.final}
\end{eqnarray}
%

\section{Calculation of $\Sigma_s^{\text{(medium)}}$}\label{appendix.sigma.s.BT}

We have from Eq.~(\ref{eq.sigma.sp.1})
\begin{eqnarray}
\Sigma_s^{\text{(medium)}} &=& -\FB{\frac{g^2_{\sigma NN}}{m_\sigma^2}} \int\frac{d^4p}{\FB{2\pi}^4}
\hat{T}\FB{p,m_N^*,m_1}\left.\frac{}{}2\pi \eta\FB{p\cdot u}\delta\FB{p^2-m_1^2}\right|_{m_1=m_N^*}
\end{eqnarray}
where $\hat{T}\FB{p,m_N^*,m_1}$ is given in Eq.~(\ref{eq.T}). 
Using Eqs.~(\ref{eq.eta}) and (\ref{eq.f.pm}), we can write the above equation as,
\begin{eqnarray}
\Sigma_s^{\text{(medium)}} &=& -\FB{\frac{g^2_{\sigma NN}}{m_\sigma^2}} \int\frac{d^3p}{\FB{2\pi}^3}\int_{-\infty}^{+\infty}dp^0
\hat{T}\FB{p^0,\vec{p},m_N^*,m_1}\FB{\frac{1}{2\omega_1}}\times \nn \\ 
&& \TB{\frac{}{}f_+\FB{\omega_1}\delta\FB{p^0-\omega_1}+f_-\FB{\omega_1}
\delta\FB{p^0+\omega_1}}_{m_1=m_N^*} \nn
\end{eqnarray}
where $\omega_1 = \sqrt{\vec{p}^2+m_1^2}$. 
Performing the $dp^0$ integration using the Dirac delta functions and noting that $\hat{T}\FB{p^0,\vec{p},m_N^*,m_1}$ 
is an even function of $p^0$, we get
\begin{eqnarray}
\Sigma_s^{\text{(medium)}} &=& -\FB{\frac{g^2_{\sigma NN}}{2m_\sigma^2}} \int\frac{d^3p}{\FB{2\pi}^3}
\hat{T}\FB{p^0=\Omega_p,\vec{p},m_N^*,m_1}\FB{\frac{1}{\omega_1}}\times 
\TB{\frac{}{}f_+\FB{\omega_1}+f_-\FB{\omega_1}}_{m_1=m_N^*}
\label{eq.sigma.s.p.4}
\end{eqnarray}
Substituting Eq.~(\ref{eq.T}) into (\ref{eq.sigma.s.p.4}) 
and performing the angular integration we get,
\begin{eqnarray}
\Sigma_s^{\text{(medium)}} &=& -\FB{\frac{g^2_{\sigma NN}}{8\pi^2m_\sigma^2}} \int_{0}^{\infty}\MB{\vec{p}}^2d\MB{\vec{p}}
\hat{B}\FB{\vec{p},m_N^*,m_1}\FB{\frac{1}{\omega_1}}\times 
\TB{\frac{}{}f_+\FB{\omega_1}+f_-\FB{\omega_1}}_{m_1=m_N^*}
\label{eq.sigma.s.p.5}
\end{eqnarray}
where,
\begin{eqnarray}
\hat{B}\FB{\vec{p},m_N^*,m_1} = 16m_N^*+\frac{32}{3}\FB{eB}^2m_N^*\MB{\vec{p}}^2\hat{A}_3 + 16\FB{2m_N^{*2}+\frac{4}{3}\MB{\vec{p}}^2}\times \nn \\
\SB{m_N^*\FB{\kappa_\text{p}B}^2+m_N^*\FB{\kappa_\text{n}B}^2+\FB{\MB{e}B}\FB{\kappa_\text{p}B}}\hat{A}_2 ~.
\end{eqnarray}

\subsection{Zero Temperature Case}

From Eq.~(\ref{eq.f.pm}) we have at $T=0$,
\begin{eqnarray}
\lim\limits_{T\rightarrow0} f_\pm\FB{\omega_1}=\Theta\FB{\pm\mu_\text{B}-\omega_1} \label{eq.f.T0}
\end{eqnarray} 
where $\mu_\text{B}$ is the baryon chemical potential of the medium. 
Substituting Eq.~(\ref{eq.f.T0}) into (\ref{eq.sigma.s.p.5}) 
we get,
\begin{eqnarray}
\Sigma_s^{\text{(medium)}} &=& -\FB{\frac{g^2_{\sigma NN}}{8\pi^2m_\sigma^2}} \int_{0}^{\infty}\MB{\vec{p}}^2d\MB{\vec{p}}
\left.\hat{B}\FB{\vec{p},m_N^*,m_1}\frac{1}{\omega_1}\Theta\FB{\mu_\text{B}-\omega_1}\right|_{m_1=m_N^*}~.
\end{eqnarray}

The the $d\MB{\vec{p}}$ integration of the above equation can be evaluated analytically using the following identities
\begin{eqnarray}
I_2\FB{\mu,m}&=&\int_{0}^{\sqrt{\mu^2-m^2}}\frac{\MB{\vec{p}}^2d\MB{\vec{p}}}{\sqrt{\MB{\vec{p}}^2+m^2}}
=\frac{1}{2}\TB{\mu\sqrt{\mu^2-m^2}+m^2\ln\SB{\frac{m}{\mu+\sqrt{\mu^2-m^2}}}} \\
I_4\FB{\mu,m}&=&\int_{0}^{\sqrt{\mu^2-m^2}}\frac{\MB{\vec{p}}^4d\MB{\vec{p}}}{\sqrt{\MB{\vec{p}}^2+m^2}}
=\frac{1}{8}\TB{\mu\FB{2\mu^2-5m^2}\sqrt{\mu^2-m^2}-3m^4\ln\SB{\frac{m}{\mu+\sqrt{\mu^2-m^2}}}}
\end{eqnarray}
and we get,
\begin{eqnarray}
\Sigma_s^\text{(medium)} &=& -\FB{\frac{2g^2_{\sigma NN}}{\pi^2m_\sigma^2}} \TB{m_N^*I_2\FB{\mu_\text{B},m_1}+\frac{2}{3}\FB{eB}^2m_N^*
\hat{A}_3I_4\FB{\mu_\text{B},m_1} \right. \nn \\ && \left. + 2\SB{m_N^*\FB{\kappa_\text{p}B}^2+m_N^*\FB{\kappa_\text{n}B}^2+\FB{\MB{e}B}
\FB{\kappa_\text{p}B}}\SB{m_N^{*2}\hat{A}_2I_2\FB{\mu_\text{B},m_1}
	+\frac{1}{3}\hat{A}_2I_4\FB{\mu_\text{B},m_1}} }_{m_1=m_N^*} ~.
\end{eqnarray}
It is now trivial to check that
\begin{eqnarray}
\left.\frac{}{}\hat{A}_2I_2\FB{\mu,m_1}\right|_{m_1=m_N^*} &=& \left.\frac{}{}2\hat{A}_3I_4\FB{\mu,m_1}\right|_{m_1=m_N^*} 
=\frac{\mu}{8m_N^{*2}\sqrt{\mu^2-m_N^{*2}}} = C_1\FB{\mu,m_N^*} ~~~ \text{(say)}  \\ 
\left.\frac{}{}\hat{A}_2I_4\FB{\mu,m_1}\right|_{m_1=m_N^*} &=& 
-\FB{\frac{3}{8}}\ln\SB{\frac{m_N^*}{\mu+\sqrt{\mu^2-m_N^{*2}}}}  = C_2\FB{\mu,m_N^*} ~~~ \text{(say)}~. 
\end{eqnarray}
So finally $\Sigma_s^\text{(medium)}$ becomes,
\begin{eqnarray}
\Sigma_s^\text{(medium)} &=& -\FB{\frac{2g^2_{\sigma NN}}{\pi^2m_\sigma^2}} \TB{m_N^*I_2\FB{\mu_\text{B},m_N^*}+\frac{1}{3}\FB{eB}^2m_N^*
	C_1\FB{\mu_\text{B},m_N^*} \right. \nn \\ && \left.
	+ 2\SB{m_N^*\FB{\kappa_\text{p}B}^2+m_N^*\FB{\kappa_\text{n}B}^2+\FB{\MB{e}B}\FB{\kappa_\text{p}B}}\SB{m_N^{*2}C_1\FB{\mu_\text{B},m_N^*}
		+\frac{1}{3}C_2\FB{\mu_\text{B},m_N^*}} } \label{eq.sigma.spn.final} ~.
\end{eqnarray}


\subsection{Finite Temperature Case}

At finite temperature, the $d\MB{\vec{p}}$ integration in Eq.~(\ref{eq.sigma.s.p.5}) can not be performed analytically. We simplify the 
expression by evaluating the derivatives with respect to $m_1^2$ explicitly. For this we use the following results
\begin{eqnarray}
\TB{\frac{f_\pm\FB{\omega_1}}{\omega_1}}_{m_1=m_N^*} &=& \frac{N^p_\pm}{\Omega_p} = \tilde{C}_1^{\pm p} ~~~\text{(say)} \\
\hat{A}_2\TB{\frac{f_\pm\FB{\omega_1}}{\omega_1}}_{m_1=m_N^*} &=& \frac{N^p_\pm}{8\Omega_p^5}
\TB{3+3\FB{1-N^p_\pm}\beta\Omega_p+\SB{1-3N^p_\pm+2\FB{N^p_\pm}^2}\beta^2\Omega_p^2}
=\tilde{C}_2^{\pm p} ~~~\text{(say)}\\ 
\hat{A}_3\TB{\frac{f_\pm\FB{\omega_1}}{\omega_1}}_{m_1=m_N^*} &=& \frac{N^p_\pm}{48\Omega_p^7}
\TB{15+15\FB{1-N^p_\pm}\beta\Omega_p+6\SB{1-3N^p_\pm+2\FB{N^p_\pm}^2}\beta^2\Omega_p^2 \right. \nn \\
&& \left.+\SB{1-7N^p_\pm+12\FB{N^p_\pm}^2-6\FB{N^p_\pm}^3}\beta^3\Omega_p^3}=\tilde{C}_3^{\pm p} ~~~\text{(say)}
\end{eqnarray}
and obtain from Eq.~(\ref{eq.sigma.s.p.5})
\begin{eqnarray}
\Sigma_s^{\text{(medium)}} &=& -\FB{\frac{2g^2_{\sigma NN}}{\pi^2m_\sigma^2}} \int_{0}^{\infty}\MB{\vec{p}}^2d\MB{\vec{p}}
\TB{m_N^*\FB{\tilde{C}_1^{+p}+\tilde{C}_1^{-p}}+\frac{2}{3}m_N^*\FB{eB}^2\MB{\vec{p}}^2
\FB{\tilde{C}_3^{+p}+\tilde{C}_3^{-p}} \right. \nn \\ 
&& \left. +2\FB{m_N^{*2}+\frac{2}{3}\MB{\vec{p}}^2}\SB{m_N^*\FB{\kappa_\text{p}B}^2+m_N^*\FB{\kappa_\text{n}B}^2+\FB{\MB{e}B}
	\FB{\kappa_\text{p}B}}
\FB{\tilde{C}_2^{+p}+\tilde{C}_2^{-p}}} \label{eq.sigma.spnt.final}
\end{eqnarray}

\end{document}